\documentclass[useAMS,usenatbib]{mn2e}

\usepackage{graphicx}
\usepackage{subfig}
\usepackage{threeparttable}
\usepackage{longtable}
\usepackage{txfonts}
\usepackage{natbib}
\usepackage{fixltx2e}
\usepackage{ownmacros}
\usepackage[T1]{fontenc}
\usepackage[compatibility = false]{caption}
\usepackage{aecompl}

\title[A multiplicity study of exoplanet host stars]{A lucky imaging multiplicity study of exoplanet host stars II\thanks{Based on observations collected at the German-Spanish Astronomical Centre, Calar Alto, Spain, operated jointly by the Max-Plank-Institut f\"ur Astronomie (MPIA), Heidelberg, and the Spanish National Commission for Astronomy.}\thanks{Based in part on observations obtained at Paranal Observatory in ESO program 095.C-0273(A)}}

\author[C. Ginski et al.]{C. Ginski$^{1,2}$\thanks{E-mail:
ginski@strw.leidenuniv.nl},  M. Mugrauer$^{2}$, M. Seeliger$^{2}$, S. Buder$^{2,3}$, R. Errmann$^{2,4}$, H. Avenhaus$^{5}$, \newauthor D. Mouillet$^{6,7}$, A.-L. Maire$^{8}$, and S. Raetz$^{9}$\\
$^{1}$Leiden Observatory, Leiden University, P.O. Box 9513, 2300 RA Leiden, The Netherlands\\
$^{2}$Astrophysikalisches Institut und Universit\"ats-Sternwarte Jena, Schillerg\"asschen 2, 07745 Jena, Germany\\
$^{3}$Max-Planck-Institut f{\"u}r Astronomie, K{\"o}nigstuhl 17, D-69117 Heidelberg, Germany\\
$^{4}$Isaac Newton Group of Telescopes, Santa Cruz de La Palma, Spain\\
$^{5}$Departamento  de  Astronom\`{i}a,  Universidad  de  Chile,  Casilla  36-D, Santiago, Chile\\
$^{6}$Universit\'{e} Grenoble Alpes, IPAG, F-38000 Grenoble, France\\
$^{7}$CNRS, IPAG, F-38000 Grenoble, France\\
$^{8}$INAF - Osservatorio Astronomico di Padova, Vicolo dell'Osservatorio 5, 35122 Padova, Italy\\
$^{9}$Scientific Support Office, Directorate of Science and Robotic Exploration, European Space Research and Technology Centre (ESA/ESTEC), \\
Keplerlaan 1, 2201 AZ Noordwijk, The Netherlands
}

\begin{document}

\date{Accepted 2015 XX. Received 2015 XX}

\pagerange{\pageref{firstpage}--\pageref{lastpage}} \pubyear{2015}

\maketitle

\label{firstpage}

\begin{abstract}
The vast majority of extrasolar planets are detected by indirect detection methods such as transit monitoring and radial velocity measurements. While these methods are very successful in detecting short-periodic planets, they are mostly blind to wide sub-stellar or even stellar companions on long orbits. 
In our study we present high resolution imaging observations of 63 exoplanet hosts carried out with the lucky imaging instrument AstraLux at the Calar Alto 2.2\,m telescope as well as with the new SPHERE high resolution adaptive optics imager at the ESO/VLT in the case of a known companion of specific interest.
Our goal is to study the influence of stellar multiplicity on the planet formation process.\\
We detected and confirmed 4 previously unknown stellar companions to the exoplanet hosts HD197037, HD217786, Kepler-21 and Kepler-68. In addition, we detected 11 new low-mass stellar companion candidates which must still be confirmed as bound companions. We also provide new astrometric and photometric data points for the recently discovered very close binary systems WASP-76 and HD\,2638. Furthermore, we show for the first time that the previously detected stellar companion to the HD185269 system is a very low mass binary.
Finally we provide precise constraints on additional companions for all observed stars in our sample.

\end{abstract}


\section{Introduction}

We live in a golden age for extrasolar planet discoveries. In the past decade several large radial velocity and transit surveys have discovered more than 1200 systems containing extrasolar planets (exoplanet.eu, as of July 2015). While these indirect detection methods have been incredibly successful, they have a few inherent biases.
In particular, while they are very sensitive to short-period companions (often in the order of days or weeks), they are blind to wide (sub-) stellar companions at several tens or hundreds of au. However, more than 50\% of all main sequence stars in the Galaxy and approximately half of all solar type stars are actually members of stellar multiple systems (\citealt{2000prpl.conf..703M}, \citealt{2010ApJS..190....1R}). 
It is thus of great interest to investigate the influence of stellar multiplicity on extrasolar planet formation and orbital evolution. \\
There have been a large number of theoretical and observational studies that investigated the influence of close and wide stellar companions on the various stages of the planet formation process. It is, for instance, believed that close stellar companions will truncate protoplanetary disks and shorten their dissipation timescale.
This has been observationally confirmed e.g. by \cite{2006ApJ...653L..57B} who found a significantly reduced number of disks in binary systems in their Spitzer survey of the young $\eta$\,Cha star cluster. Other studies such as \cite{2012ApJ...745...19K} find that this effect is dependent on the binary separation with significant drops of disk occurrences only observed for systems with separations smaller than $\sim$40\,au.\\
In addition to the initial conditions and timescales in the protoplanetary disk, stellar companions might also influence the accretion of planetesimals by exciting higher eccentricities and velocities which might lead to more destructive collisions (see e.g. \citealt{2007arXiv0705.3421K} or \citealt{2008MNRAS.386..973P}). However, recent studies find that this effect might be mitigated by the gravitational force of sufficiently massive disks (\citealt{2013ApJ...764L..16R}).\\
Finally, stellar companions might have a major influence on the observed semi-major axis, inclination and eccentricity distributions of extrasolar planets. Studies by \cite{2007ApJ...669.1298F} and \cite{2015ApJ...799...27P} suggest that Kozai-Lidov type interactions between planets and stellar companions, in combination with tidal friction, might explain some of the observed extreme short period orbits. Other studies (e.g. \citealt{2011Natur.473..187N}) suggest that such interactions could explain very eccentric planet orbits or spin-orbit misalignment. 
For a comprehensive overview of all these effects we suggest the article by \cite{2014arXiv1406.1357T}.\\
To study these effects, it is necessary to find the true fraction of multiple stellar systems amongst extrasolar planet host stars. Diffraction- or seeing-limited imaging is a primary tool for this purpose, in particular to find multiple stellar systems with planets in S-type oribts, i.e. the planets orbit one of the stellar components of the system. This orbit configuration accounts for the majority of multiple stellar exoplanet systems (see e.g. \citealt{2012A&A...542A..92R}). \\
There have been a number of imaging studies in the past such as \cite{2007A&A...474..273E}, \cite{2007AandA...469..755M}, \cite{2009A&A...498..567D}, \cite{2011A&A...528A...8C}, \cite{2012A&A...546A..10L}, or more recently \cite{2014AJ....148...78D}, \cite{2014MNRAS.439.1063M}, \cite{2015MNRAS.450.3127M} and \cite{2015A&A...575A..23W}.\\
In this work we present the results of our ongoing multiplicity study employing the lucky imaging instrument AstraLux (\citealt{2008SPIE.7014E..48H}) at the Calar Alto 2.2\,m telescope. In particular we present results for 63 systems obtained between 2011 and 2015. Results prior to that can be found in the first publication of our survey in \cite{2012MNRAS.421.2498G}. Our targets are stars around which an exoplanet has been detected by radial velocity or transit observations and which have not yet been observed with high resolution imaging. 
We further limit our sample to stars within $\sim$200\,pc (with few exceptions) so that we are able to confirm detected companion candidates via common proper motion analysis. In addition to our lucky imaging observations, we complement our study with extreme adaptive optics supported images from the new planet hunting instrument SPHERE (\citealt{2008SPIE.7014E..18B}) at the ESO/VLT. \\
We derive astrometric and photometric data of all detected companion candidates and perform common proper motion analysis for all systems with more than one observation epoch. Finally we provide detailed detection limits on all observed systems. 

\section{Observations and data reduction}

\begin{figure*}
\subfloat[HD\,2638]{
\includegraphics[scale=0.25]{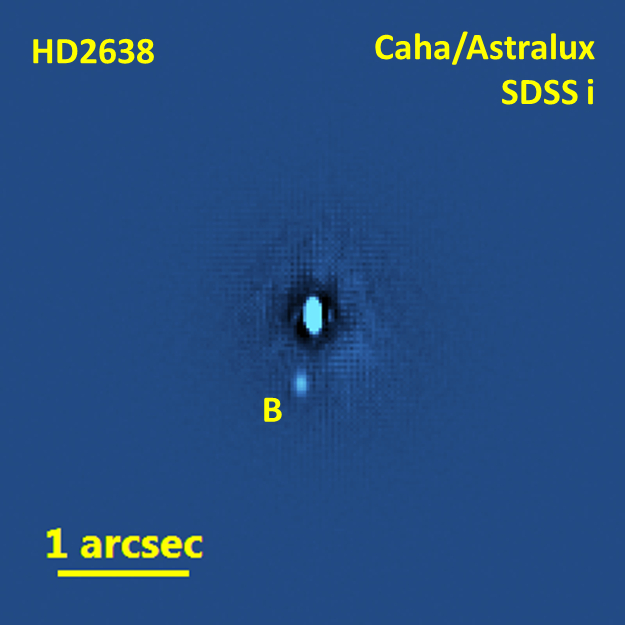}
\label{HD2638}
}
\subfloat[HAT-P-7]{
\includegraphics[scale=0.25]{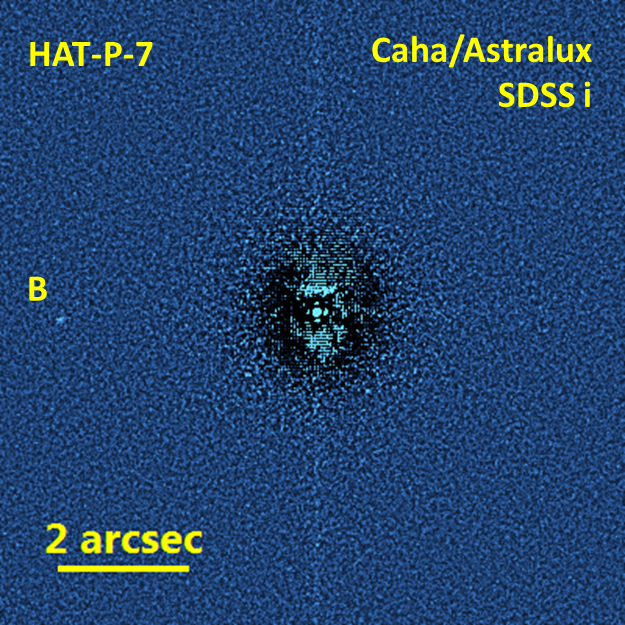}
\label{hat-p-7}
}
\subfloat[HD\,185269]{
\includegraphics[scale=0.2545]{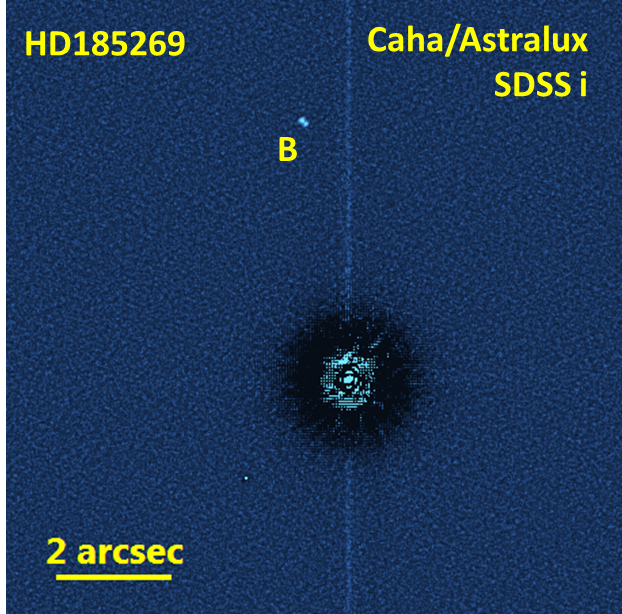}
\label{hd185269-astralux}
}

\subfloat[WASP-76]{
\includegraphics[scale=0.25]{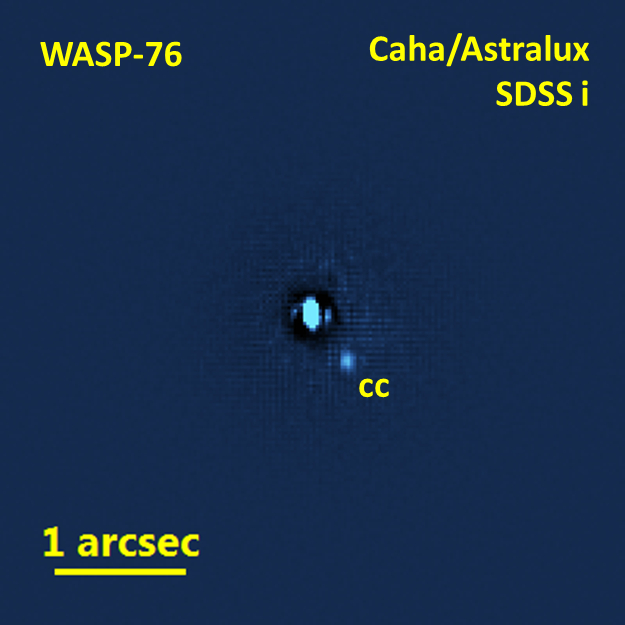}
\label{wasp-76}
}
\subfloat[HAT-P-32]{
\includegraphics[scale=0.25]{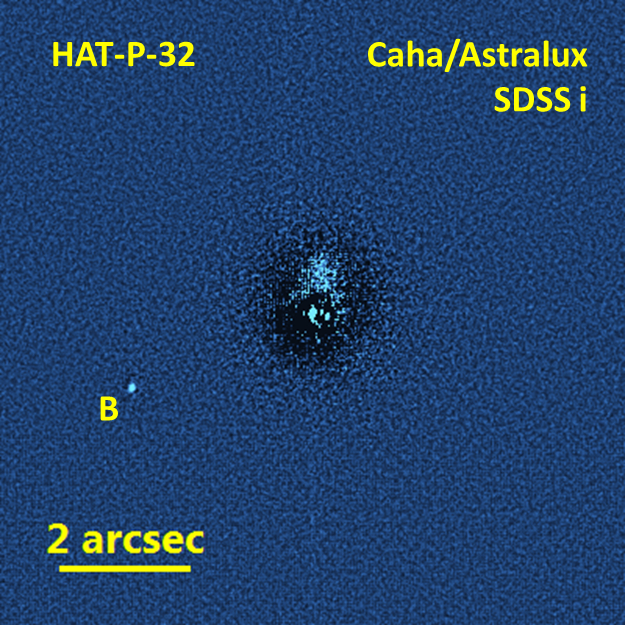}
\label{hat-p-32}
}

\caption[]{Images of known low-mass stellar companions to exoplanet host stars, followed up in our multiplicity study. The halos of the bright host stars were removed by high-pass filtering. North is up and East is to the left.} 
\label{known-comp-images}
\end{figure*}

\begin{figure*}
\subfloat[HD\,10697]{
\includegraphics[scale=0.2]{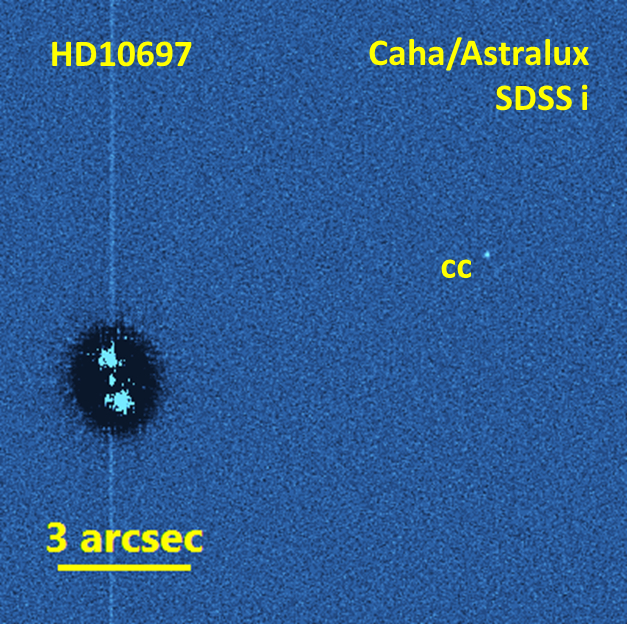}
\label{hd10697}
}
\subfloat[HD\,43691]{
\includegraphics[scale=0.2]{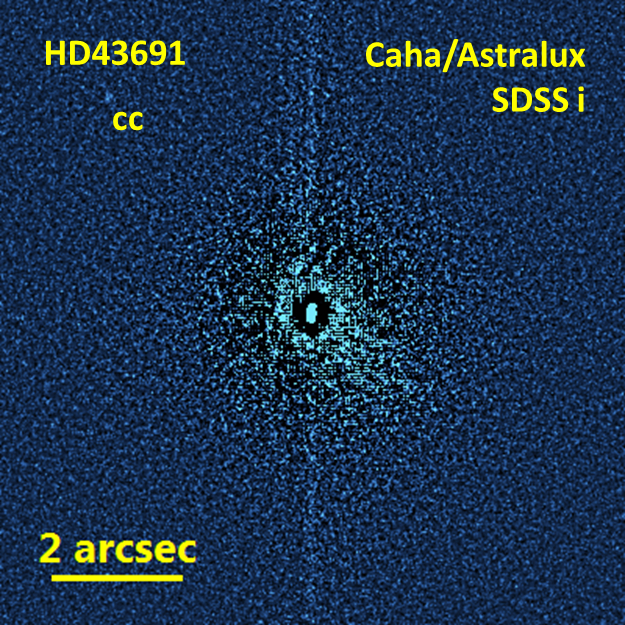}
\label{hd43691}
}
\subfloat[HD\,116029]{
\includegraphics[scale=0.2]{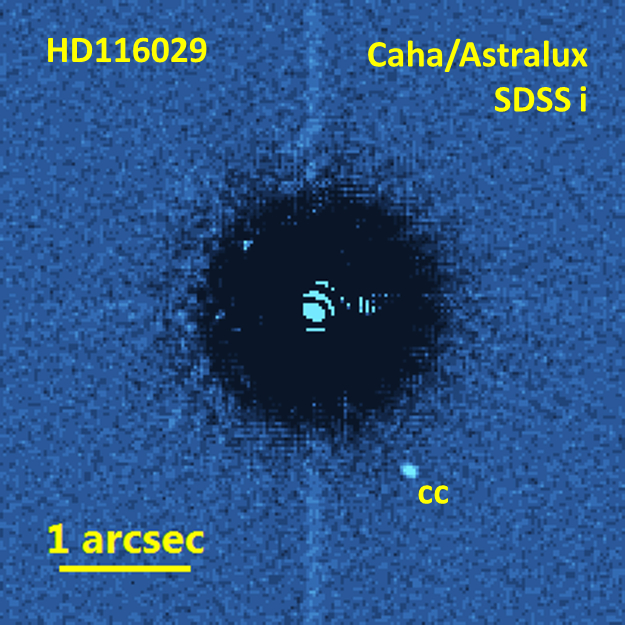}
\label{hd116029}
}
\subfloat[HAT-P-18]{
\includegraphics[scale=0.2]{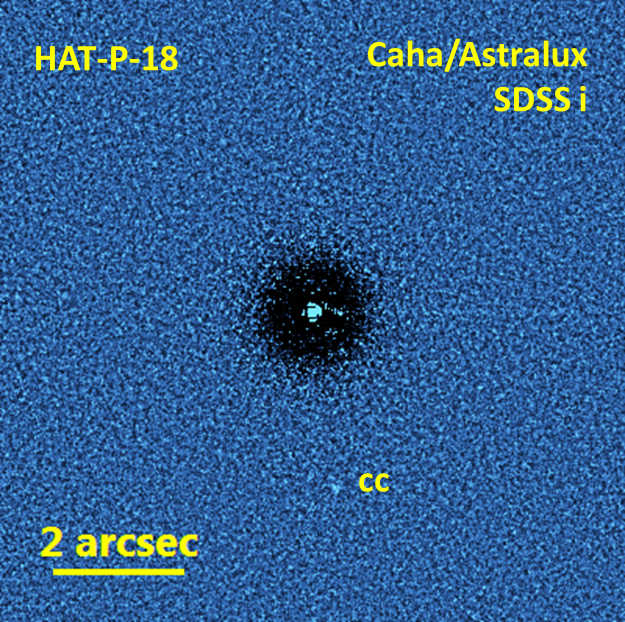}
\label{hat-p-18}
}

\subfloat[Kepler-37]{
\includegraphics[scale=0.2]{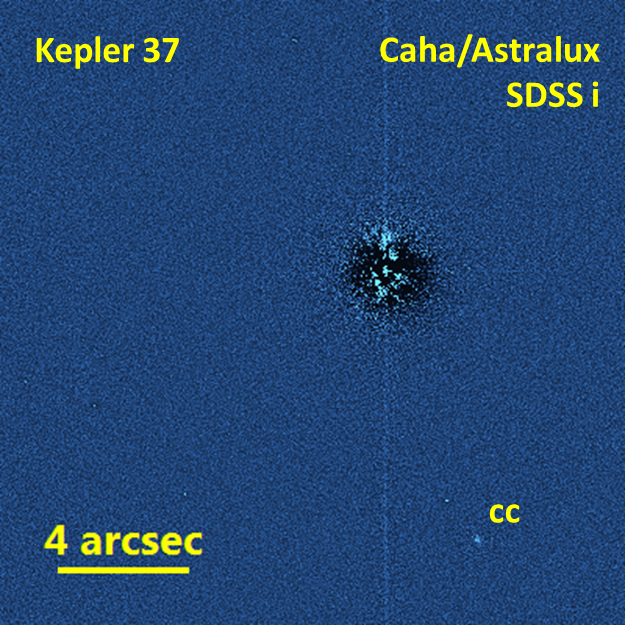}
\label{kepler37}
}
\subfloat[Kepler-21]{
\includegraphics[scale=0.2]{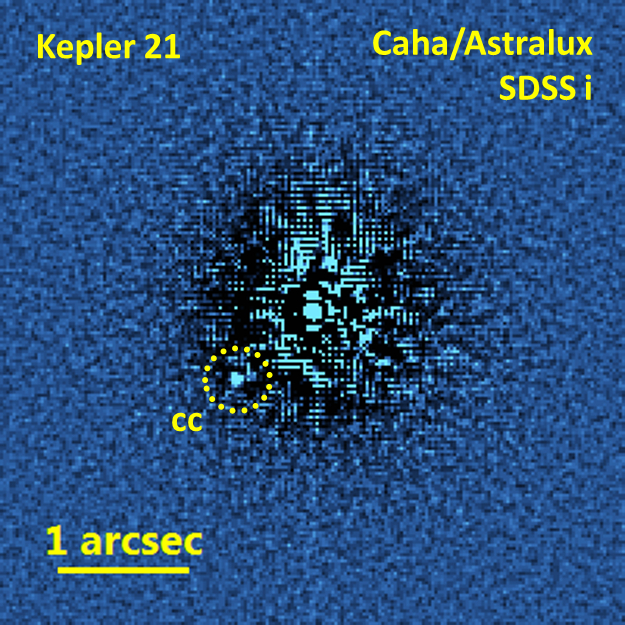}
\label{kepler21}
}
\subfloat[Kepler-68]{
\includegraphics[scale=0.199]{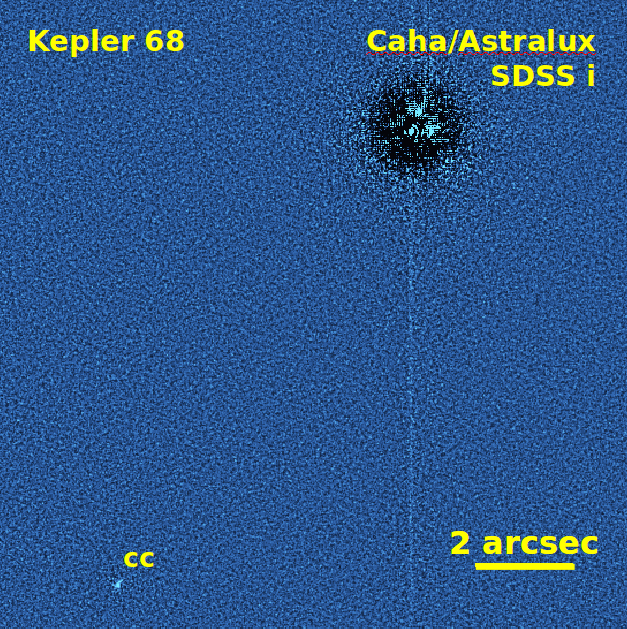}
\label{kepler68}
}
\subfloat[Kepler-42]{
\includegraphics[scale=0.2]{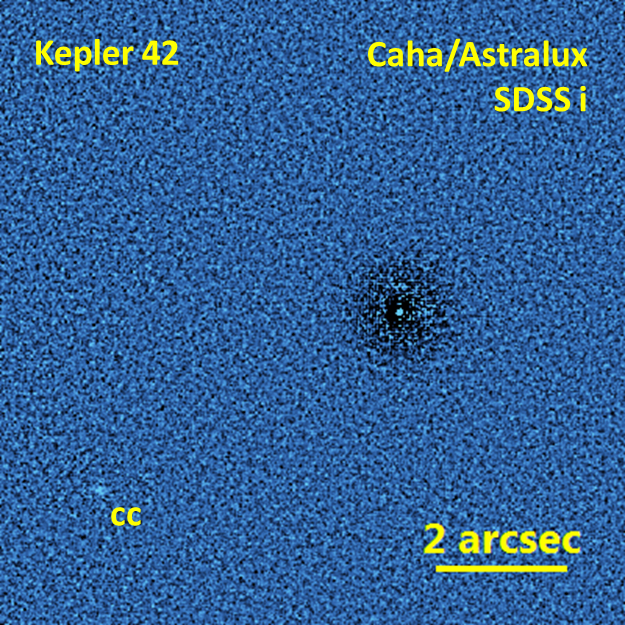}
\label{kepler42}
}

\subfloat[HD\,188015]{
\includegraphics[scale=0.2]{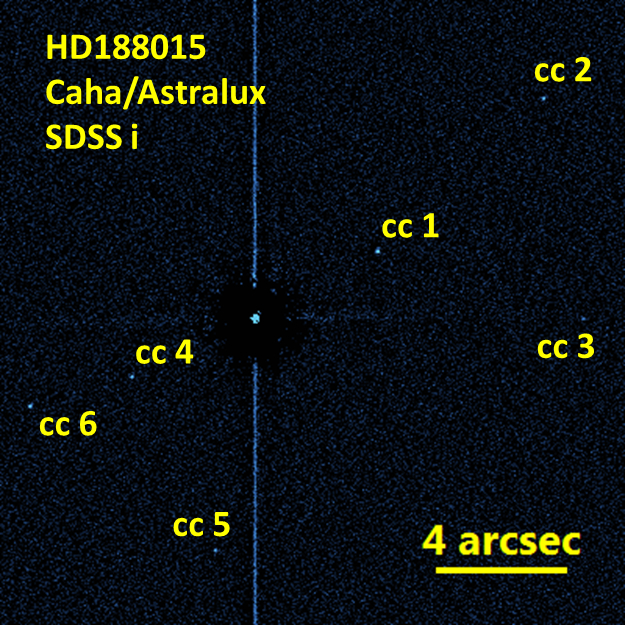}
\label{hd188015}
}
\subfloat[HD\,197037]{
\includegraphics[scale=0.2]{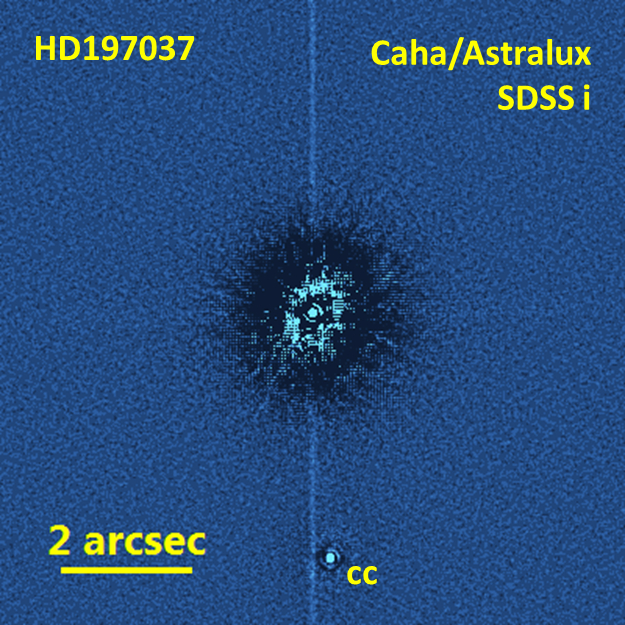}
\label{hd197037}
}
\subfloat[HD\,217786]{
\includegraphics[scale=0.2]{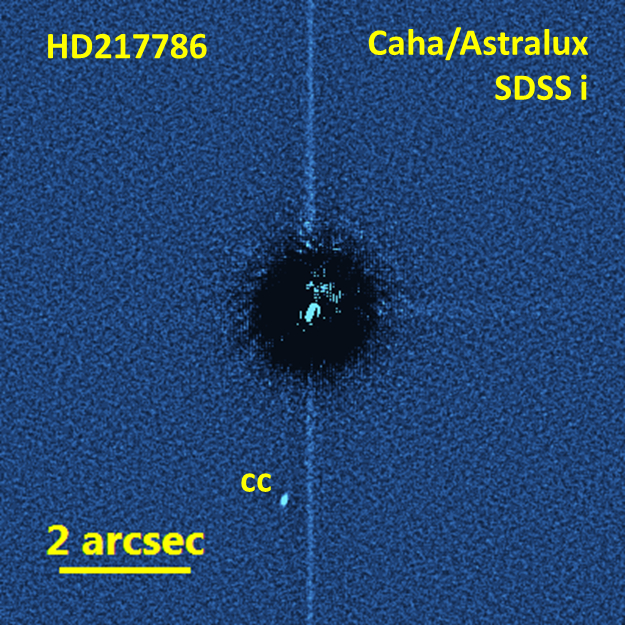}
\label{hd217786}
}

\caption[]{Images of all newly detected companion candidates during the course of our multiplicity study with Astralux at the Calar Alto 2.2\,m telescope. Spatial scaling of each image is indicated. The companion candidates (cc) are marked in all images. All images were high-pass filtered to remove the bright halo of the host star. North is always up and East is to the left.}
\label{new-comp-images}
\end{figure*}

\begin{figure}

\includegraphics[scale=0.275]{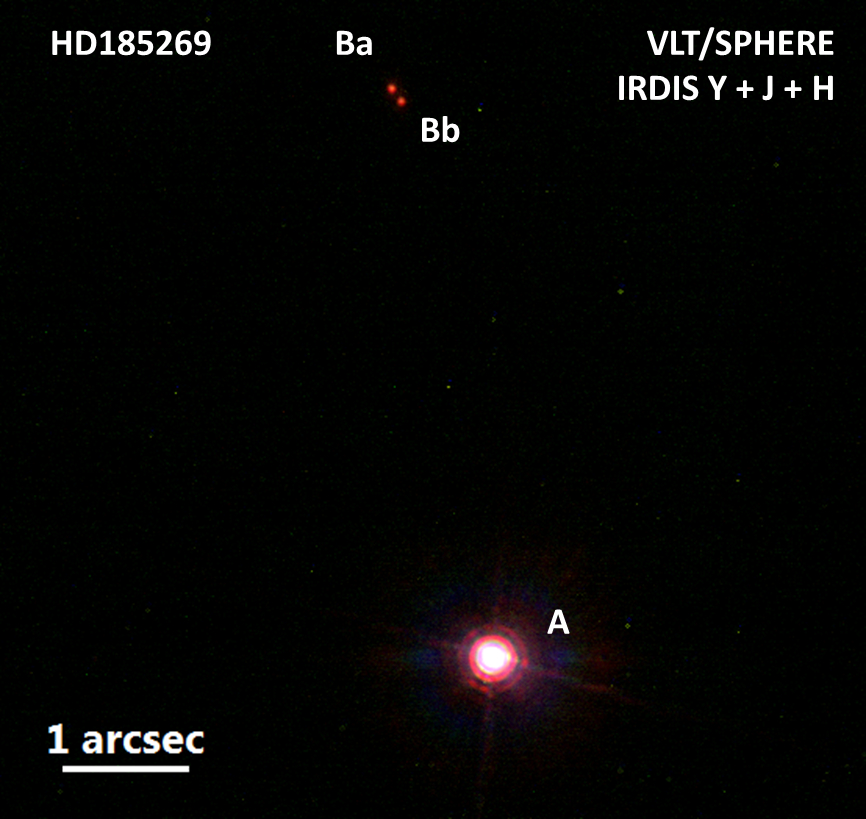}

\caption[]{Composite color image of the exoplanet host star HD\,185269 and its companion taken with SPHERE/IRDIS on 02-05-2015. Red, green and blue channels are H, J and Y-band data respectively. In the SPHERE/IRDIS image the low mass stellar companion discovered by us (\citealt{2012MNRAS.421.2498G}) is for the first time resolved as a low mass stellar binary. North is up and East is to the left.} 
\label{hd185269-sphere}
\end{figure}

The observations presented in this study were undertaken between July 2011 and March 2015 with the lucky imaging instrument AstraLux at the Calar Alto Observatory. 
In addition, we present data for one system which was taken with the new SPHERE planet hunting instrument at the ESO VLT during guaranteed time observations (GTO) in May 2015. \\
For our lucky imaging observations we used short exposures times in the same order as the coherence time of the atmosphere (e.g. \cite{2008SPIE.7014E..48H} measure a speckle coherence times at the Calar Alto of 36\,ms). 
We then recorded a large number of individual images (typically 50000) of which we only used subsets with the highest Strehl ratio (\citealt{Strehl1902}) for final combination. 
The lucky imaging technique is described in detail in e.g. \cite{2006AA...446..739L}. All lucky imaging observations were undertaken using the SDSS i filter. 
The electron multiplying gain of the instrument was adjusted individually for each target to enable high signal-to-noise without saturating the primary star. 
We also adjusted the focus of the instrument several times during the night to ensure highest image quality. 
In our 2011, 2013 and 2014 observations in visitor mode, we used the full field of view of the detector of 24\,$\times$\,24 arcsec with the shortest possible exposure time of 29.54\,ms in frame transfer mode. 
For the brightest targets we used shorter integrations times without frame transfer mode and less overall frames due to larger overheads, i.e. significantly increased readout time. 
In the 2015 observations in service mode the instrument was used in windowed mode, reading only half of the field of view. This enabled shorter exposure times of typically 15.03\,ms. Details for each system are given in Tab.~\ref{tab:obs-overview}.\\
Data reduction of the lucky imaging data included flat fielding with sky flats taken during dawn, as well as bias subtraction. Bias frames were taken before each science exposure with the same gain settings as the science target. 
After flat fielding and bias subtraction, the Strehl ratio in each image was measured and then only the images with the 10\%, 5\% and 1\% best Strehl ratios were aligned and combined respectively\footnote{If not otherwise stated we generally used the best 10\% images for subsequent analysis.}. 
For the final data reduction we utilized the native AstraLux pipeline available at Calar Alto (described in detail by \citealt{2008SPIE.7014E..48H}), as well as our own pipeline for the reduction of lucky imaging data. Our own pipeline was used in all those (few) cases where the Calar Alto pipeline produced no output due to software malfunction.
Final images with detected known companions as well as new companion candidates are shown in Fig.~\ref{known-comp-images} and \ref{new-comp-images}. We show the 2013 data when available, since it is in general of slightly higher quality than the 2014 data due to better weather conditions (higher coherence time, no clouds). 
To enhance the contrast between the bright primary stars and the faint companion candidates, we have employed high pass filtering on the images. \\
In addition, we did use SPHERE's near infrared camera IRDIS (\citealt{2008SPIE.7014E..3LD}) in dual band imaging mode (\citealt{2010MNRAS.407...71V}) to image the HD\,185269 system in Y, J and H-band with broad band filters on 02-05-2015. The specific interest in this system was triggered by an observed elongation of the companion's PSF in our AstraLux observations. 
We used the minimal exposure time of 0.84\,s without coronagraph and with neutral density filter, which led to only minor saturation of the core of the primary star's point spread function (PSF) in Y and H-band, and no saturation in J-band. For each filter setting we took a total of 20 individual exposures for a total integration time of 16.8\,s.
All individual images in each band were median combined and then flat fielded and dark subtracted. Since we did not apply a dither pattern in this very short observation sequence, we then used a bad pixel mask (created from flat and dark frames) to eliminate bad pixels. 
Finally we combined both images of the dual imaging mode in each band. A resulting combined color image is shown in Fig.~\ref{hd185269-sphere}.

\begin{table*}
%\begin{longtable*}
 %\centering
 %\begin{minipage}{200mm}
  \caption{Observation summary of all targets observed with AstraLux at the Calar Alto 2.2\,m telescope. We give the total integration time for each target for a frame selection rate of 10\,\%.}
  \begin{tabular}{@{}lccccccc@{}}
  \hline
        
 Star 	& R.A.			& DEC			& epoch 	& \# of frames		& exposure time [ms] & tot. integ. time [s] & field of view [arcsec]		\\
\hline
HD2638	&	00 29 59.87274	&	-05 45 50.4009	&	19-08-2014	&	50000	&	29.54	&	147.70	&	24$\times$24	\\
HD2952	&	00 33 10.39467	&	+54 53 41.9440	&	19-08-2014	&	50000	&	29.54	&	147.70	&	24$\times$24	\\
HD5608	&	00 58 14.21893	&	+33 57 03.1843	&	17-01-2013	&	50000	&	29.54	&	147.70	&	24$\times$24	\\
HD5891	&	01 00 33.19204	&	+20 17 32.9381	&	17-01-2013	&	50000	&	29.54	&	147.70	&	24$\times$24	\\
HD8574	&	01 25 12.51565	&	+28 34 00.1010	&	30-06-2013	&	50000	&	29.54	&	147.70	&	24$\times$24	\\
HD10697	&	01 44 55.82484	&	+20 04 59.3381	&	20-08-2014	&	50000	&	29.54	&	147.70	&	24$\times$24	\\
WASP-76	&	01 46 31.8590	&	+02 42 02.065	&	19-08-2014	&	50000	&	29.54	&	147.70	&	24$\times$24	\\
HAT-P-32	&	02 04 10.278	&	+46 41 16.21	&	19-08-2014	&	60000	&	29.54	&	177.24	&	24$\times$24	\\
HD12661	&	02 04 34.28834	&	+25 24 51.5031	&	20-08-2014	&	50000	&	29.54	&	147.70	&	24$\times$24	\\
HD13189	&	02 09 40.17260	&	+32 18 59.1649	&	20-08-2014	&	50000	&	29.54	&	147.70	&	24$\times$24	\\
HD13908	&	02 18 14.56056	&	+65 35 39.6988	&	19-08-2014	&	50000	&	29.54	&	147.70	&	24$\times$24	\\
HD15779	&	02 32 09.42200	&	-01 02 05.6236	&	17-01-2013	&	50000	&	29.54	&	147.70	&	24$\times$24	\\
HD285507	&	04 07 01.22653	&	+15 20 06.0989	&	20-08-2014	&	50000	&	29.54	&	147.70	&	24$\times$24	\\
HD290327	&	05 23 21.56490	&	-02 16 39.4302	&	10-03-2015	&	50000	&	15.03	&	75.15	&	12$\times$12	\\
HD40979	&	06 04 29.94214	&	+44 15 37.5940	&	10-03-2015	&	50000	&	29.54	&	147.70	&	12$\times$12	\\
HD43691	&	06 19 34.67623	&	+41 05 32.3113	&	10-03-2015	&	16383	&	15.01	&	24.59	&	12$\times$12	\\
HD45350	&	06 28 45.71155	&	+38 57 46.6670	&	10-03-2015	&	50000	&	15.01	&	75.05	&	12$\times$12	\\
Omi Uma	&	08 30 15.87064	&	+60 43 05.4115	&	10-03-2015	&	20000	&	5.01	&	10.02	&	12$\times$12	\\
GJ328	&	08 55 07.597	&	+01 32 56.44	&	10-03-2015	&	50000	&	15.01	&	75.05	&	12$\times$12	\\
HD95089	&	10 58 47.73629	&	+01 43 45.1758	&	10-03-2015	&	32766	&	15.01	&	49.18	&	12$\times$12	\\
HD96063	&	11 04 44.45463	&	-02 30 47.5867	&	10-03-2015	&	50000	&	15.01	&	75.05	&	12$\times$12	\\
HD99706	&	11 28 30.21370	&	+43 57 59.6902	&	10-03-2015	&	50000	&	15.01	&	75.05	&	12$\times$12	\\
HD100655	&	11 35 03.75349	&	+20 26 29.5713	&	10-03-2015	&	50000	&	15.01	&	75.05	&	12$\times$12	\\
HIP57274	&	11 44 40.96488	&	+30 57 33.4552	&	10-03-2015	&	50000	&	15.01	&	75.05	&	12$\times$12	\\
HD102329	&	11 46 46.64518	&	+03 28 27.4563	&	10-03-2015	&	50000	&	15.01	&	75.05	&	12$\times$12	\\
HD106270	&	12 13 37.28529	&	-09 30 48.1691	&	10-03-2015	&	16383	&	15.01	&	24.59	&	12$\times$12	\\
HD113337	&	13 01 46.92669	&	+63 36 36.8092	&	10-03-2015	&	50000	&	15.01	&	75.05	&	12$\times$12	\\
HD116029	&	13 20 39.54263	&	+24 38 55.3080	&	30-06-2013	&	50000	&	29.54	&	147.70	&	24$\times$24	\\
	&	13 20 39.54263	&	+24 38 55.3080	&	20-08-2014	&	60000	&	29.54	&	177.24	&	24$\times$24	\\
HD120084	&	13 42 39.20186	&	+78 03 51.9756	&	10-03-2015	&	50000	&	15.01	&	75.05	&	12$\times$12	\\
Beta UMi	&	14 50 42.32580	&	+74 09 19.8142	&	10-03-2015	&	20000	&	4	&	8.00	&	12$\times$12	\\
HD131496	&	14 53 23.02871	&	+18 14 07.4562	&	30-06-2013	&	50000	&	29.54	&	147.70	&	24$\times$24	\\
HD131496	&	14 53 23.02871	&	+18 14 07.4562	&	10-03-2015	&	50000	&	15.01	&	75.05	&	12$\times$12	\\
HD136726	&	15 17 05.88899	&	+71 49 26.0466	&	30-06-2013	&	50000	&	29.54	&	147.70	&	24$\times$24	\\
11 UMi	&	15 17 05.88899	&	+71 49 26.0466	&	10-03-2015	&	50000	&	15.01	&	75.05	&	12$\times$12	\\
HD136512	&	15 20 08.55879	&	+29 36 58.3488	&	01-07-2013	&	50000	&	29.54	&	147.70	&	24$\times$24	\\
Omi CrB	&	15 20 08.55879	&	+29 36 58.3488	&	10-03-2015	&	50000	&	15.01	&	75.05	&	12$\times$12	\\
HD139357	&	15 35 16.19886	&	+53 55 19.7129	&	01-07-2013	&	50000	&	29.54	&	147.70	&	24$\times$24	\\
HD145457	&	16 10 03.91431	&	+26 44 33.8927	&	01-07-2013	&	50000	&	29.54	&	147.70	&	24$\times$24	\\
HD152581	&	16 53 43.58257	&	+11 58 25.4822	&	01-07-2013	&	50000	&	29.54	&	147.70	&	24$\times$24	\\
HAT-P-18	&	17 05 23.151	&	+33 00 44.97	&	30-06-2013	&	50000	&	29.54	&	147.70	&	24$\times$24	\\
	&	17 05 23.151	&	+33 00 44.97	&	19-08-2014	&	65540	&	29.54	&	193.61	&	24$\times$24	\\
	&	17 05 23.151	&	+33 00 44.97	&	20-08-2014	&	50000	&	29.54	&	147.70	&	24$\times$24	\\
HD156279	&	17 12 23.20383	&	+63 21 07.5391	&	01-07-2013	&	50000	&	29.54	&	147.70	&	24$\times$24	\\
HD163607	&	17 53 40.49479	&	+56 23 31.0417	&	30-06-2013	&	50000	&	29.54	&	147.70	&	24$\times$24	\\
HD163917	&	17 59 01.59191	&	-09 46 25.0798	&	30-06-2013	&	50000	&	29.54	&	147.70	&	24$\times$24	\\
HIP91258	&	18 36 53.15422	&	+61 42 09.0124	&	20-08-2014	&	50000	&	29.54	&	147.70	&	24$\times$24	\\
Kepler-37	&	18 56 14.3063	&	+44 31 05.356	&	19-08-2014	&	50000	&	29.54	&	147.70	&	24$\times$24	\\
Kepler-21	&	19 09 26.83535	&	+38 42 50.4593	&	01-07-2013	&	50000	&	29.54	&	147.70	&	24$\times$24	\\
	&	19 09 26.83535	&	+38 42 50.4593	&	20-08-2014	&	50000	&	29.54	&	147.70	&	24$\times$24	\\
HD180314	&	19 14 50.20890	&	+31 51 37.2569	&	30-06-2013	&	50000	&	29.54	&	147.70	&	24$\times$24	\\
Kepler-63	&	19 16 54.294	&	+49 32 53.51	&	20-08-2014	&	50000	&	29.54	&	147.70	&	24$\times$24	\\
Kepler-68	&	19 24 07.7644	&	+49 02 24.957	&	01-07-2013	&	50000	&	29.54	&	147.70	&	24$\times$24	\\
	&	19 24 07.7644	&	+49 02 24.957	&	19-08-2014	&	50000	&	29.54	&	147.70	&	24$\times$24	\\
Kepler-42	&	19 28 52.556	&	+44 37 09.62	&	30-06-2013	&	50000	&	29.54	&	147.70	&	24$\times$24	\\
HAT-P-7	&	19 28 59.3616	&	+47 58 10.264	&	19-08-2014	&	50000	&	29.54	&	147.70	&	24$\times$24	\\
HD185269	&	19 37 11.74092 	&	+28 29 59.5055	&	30-06-2013	&	50000	&	29.54	&	147.70	&	24$\times$24	\\
	&	19 37 11.74092 	&	+28 29 59.5055	&	19-08-2014	&	50000	&	29.54	&	147.70	&	24$\times$24	\\
HD188015	&	19 52 04.54338	&	+28 06 01.3517	&	30-06-2013	&	50000	&	29.54	&	147.70	&	24$\times$24	\\
	&	19 52 04.54338	&	+28 06 01.3517	&	20-08-2014	&	50000	&	29.54	&	147.70	&	24$\times$24	\\
HD190360	&	20 03 37.40587	&	+29 53 48.4944	&	01-07-2013	&	50000	&	29.54	&	147.70	&	24$\times$24	\\

\hline\end{tabular}
%\label{tab:obs-overview}
%\end{minipage}
\end{table*}
%\end{longtable*}	

\begin{table*}
\ContinuedFloat
%\phantomcaption

%\begin{longtable*}
 %\centering
 %\begin{minipage}{200mm}
  \caption{\textbf{Continued}}
  \begin{tabular}{@{}lccccccc@{}}
  \hline
        
 Star 				& R.A.							& DEC				& epoch & \# of frames		& exposure time [ms] & tot. integ. time [s] & field of view [arcsec]								\\
\hline

HD197037	&	20 39 32.96014	&	+42 14 54.7845	&	01-07-2013	&	50000	&	29.54	&	147.70	&	24$\times$24	\\
	&	20 39 32.96014	&	+42 14 54.7845	&	19-08-2014	&	50000	&	29.54	&	147.70	&	24$\times$24	\\
HD206610	&	21 43 24.90004	&	-07 24 29.7086	&	20-08-2014	&	50000	&	29.54	&	147.70	&	24$\times$24	\\
HD208527	&	21 56 23.98467	&	+21 14 23.4961	&	20-08-2014	&	50000	&	29.54	&	147.70	&	24$\times$24	\\
HD210277	&	22 09 29.86552	&	-07 32 55.1548	&	19-08-2014	&	50000	&	29.54	&	147.70	&	24$\times$24	\\
HD217786	&	23 03 08.205	&	-00 25 46.66	&	28-07-2011	&	50000	&	29.54	&	147.70	&	24$\times$24	\\
	&	23 03 08.205	&	-00 25 46.66	&	30-06-2013	&	50000	&	29.54	&	147.70	&	24$\times$24	\\
	&	23 03 08.205	&	-00 25 46.66	&	20-08-2014	&	50000	&	29.54	&	147.70	&	24$\times$24	\\
HD240210	&	23 10 29.2303	&	+57 01 46.035	&	01-07-2013	&	50000	&	29.54	&	147.70	&	24$\times$24	\\
HD219828	&	23 18 46.73445	&	+18 38 44.6021	&	30-06-2013	&	19214	&	29.54	&	56.76	&	24$\times$24	\\
HD220074	&	23 20 14.37962	&	+61 58 12.4578	&	19-08-2014	&	50000	&	29.54	&	147.70	&	24$\times$24	\\
HD222155	&	23 38 00.30741	&	+48 59 47.4907	&	01-07-2013	&	50000	&	29.54	&	147.70	&	24$\times$24	\\

\hline\end{tabular}
\label{tab:obs-overview}
%\end{minipage}
\end{table*}
%\end{longtable*}

\section{Astrometric calibration and measurements}
\label{sec:astrometry}
\begin{figure}
\subfloat[HD\,43691\,A]{
\includegraphics[scale=0.25]{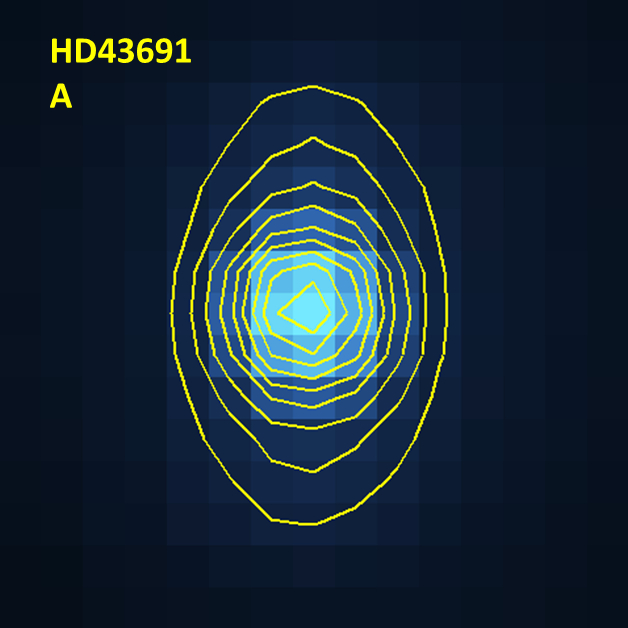}
\label{hd43691-contour-A}
}

\subfloat[HD\,43691\,cc]{
\includegraphics[scale=0.25]{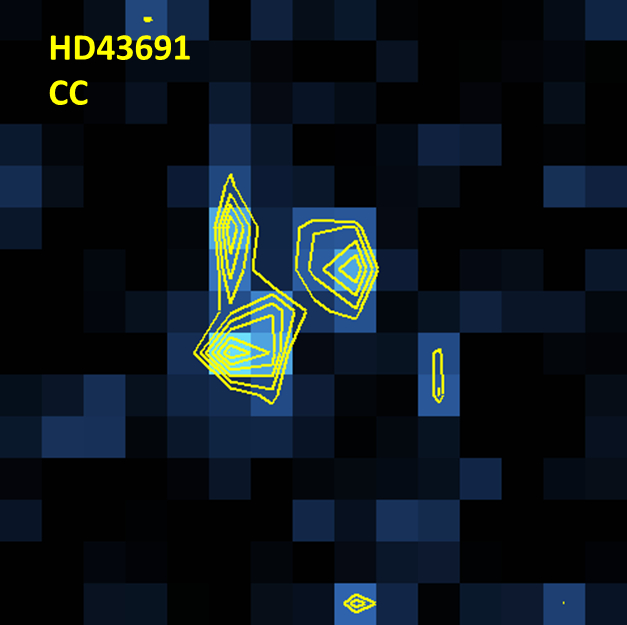}
\label{hd43691-contour-cc}
}

\caption[]{Close up of the primary star and companion candidate PSF of HD\,43691. The primary star appears relatively circular with a halo that extends in the north-south direction. 
The companion candidate shows at least two distinct brightness peaks that are extending at an angle of approximately 135\,deg. Contour lines have been overplotted to guide the eye. }
\label{hd43691-contour}
\end{figure}

The most reliable method to determine if individual companion candidates are bound to the systems around which they are discovered is to ascertain if they exhibit the same proper motion as the primary star of the system. 
For this purpose we are measuring the separation and relative position angle (PA) of all newly discovered companion candidates relative to the primary star. To ensure that our astrometric measurements can be compared between different observation epochs as well as with measurements done with different instrument, we took astrometric calibration images in each observation epoch. 
In 2013 and 2014 we used the center of the globular cluster M\,15 for this purpose. In the 2015 observation epoch M\,15 was not visible and we imaged three wide binary systems instead (HIP\,72508, HI\,P80953 and HIP\,59585). To calibrate the pixel scale as well as the orientation of the detector we used as reference HST observations of M 15 that were taken on 22-10-2011 with the Wide Field Camera 3 (WFC3, \citealt{2008SPIE.7010E..1EK}). 
In the case of the binary stars, we used all measurements of the respective systems in the Washington Double Star Catalog (\citealt{2001AJ....122.3466M}) as reference. We applied a linear fit to these available measurements to correct for the slow orbital motion of these wide binaries.
For the calibration using cluster data, we measured individual star positions in our AstraLux image and the HST reference image with IDL\footnote{Interactive Data Language} starfinder (\citealt{2000SPIE.4007..879D}), which fits a reference PSF to each star position. The reference PSF was created from the data itself. We then used our own cross correlation routines to identify the same stars in both images. Finally we calculated separations and relative orientations of each star relative to all other stars. This was done for 92 stars in 2013 and 90 stars in 2014.
We then used the known astrometric calibration of the HST reference image to calculate an astrometric solution for each individual measurement. To exclude stars with a strong proper motion or possibly misidentified stars we employed sigma clipping. The final astrometric solution for the 2013 and 2014 observations is the median of all computed solutions. 
We give the results in Tab.~\ref{tab: astrocal}. The listed uncertainties are the standard deviations of all astrometric solutions.\\
In the case of the binary stars, we only have two objects in the field of view, thus we could not create a reference PSF from the data. Instead we are fitting a two dimensional Gaussian to the star positions. We checked that this approach is valid by comparing similar measurements in the cluster images with the starfinder results. The deviations between the two methods were typically much smaller than the measurement accuracy.
The result of the binary calibration is also given in Tab.~\ref{tab: astrocal}. We used the weighted average of the three solutions calculated from the individual binary systems. For the uncertainty, we conservatively assumed the largest individual uncertainty that we measured. The uncertainty of the calibration includes the uncertainty of the linear orbital motion fit mentioned earlier. We note that calibrations using binary stars are prone to systematic offsets due to unaccounted for (or underestimated) orbital motion of the systems. We thus caution that the result of the 2015 calibration might still suffer from such an offset.\\
We have one companion candidate which was already observed in July of 2011 for the first time. In this case we utilized the astrometric calibration derived by us with cluster and binary data in \cite{2012MNRAS.421.2498G}.\\
For the SPHERE/IRDIS data we used the astrometric solution calculated by the SPHERE consortium for the GTO run in which the data was taken. This astrometric calibration was derived from multiple observations of the globular clusters 47\,Tuc and NGC\,6380, for which also precise HST reference observations as well as proper motions for individual cluster members are available. There is a small dependence of the pixel scale on the utilized filter; for our Y-band observations we used 12.234$\pm$0.029 mas/pix and -1.78$\pm$0.13 deg, while we used 12.214$\pm$0.029 mas/pix for the J-band, and 12.210$\pm$0.029 mas/pix for the H-band (the detector orientation is not influenced by the filter choice). In addition, IRDIS shows a small anamorphism between the detector x and y direction.  
This was also determined from observations of the globular cluster 47\,Tuc. To correct for this anamorphism we multiplied the separation in y by a factor of 1.0062. A detailed description of the IRDIS astrometric calibration is given in \cite{2015arXiv151104072M}.\\
The measurements of the relative positions of companion candidates to the primary stars was also done by fitting a 2 dimensional Gaussian to both objects since there were no other objects in the field of view to build a reference PSF. Also, it is problematic to build an average reference PSF from different data sets, since the shape of the PSF will highly depend on the atmospheric conditions and the height of the target above the horizon.
To ensure that we obtained a stable fitting result, we repeated the fitting procedure for each object at least 20 times with slightly different starting positions and fitting box sizes. For companion candidates that were separated by less than 2\,arcsec from the bright primary stars, we removed the primary stars' bright halo by high pass filtering before we measured the companion candidates position. All results are listed in Tab.~\ref{tab:measurements}. The given uncertainties are the uncertainties of the Gaussian fitting added in quadrature to the uncertainties of the astrometric calibration. Multiple observation epochs were available for several systems. We discuss these systems in the following in detail and test if the companion candidates are co-moving with the primary stars.\\
\\
\textit{WASP-76}\\
\\
WASP-76 was observed by us only once in August of 2014. We detected a faint companion candidate $\sim$0.44\,arcsec to the south-west of the star. Two months later in October of 2014, the target was observed also with AstraLux by \cite{2015AandA...579A.129W}, who also detected this companion candidate and claim that it is likely a bound companion due to the decreasing likelihood of background objects with decreasing separation. We used their discovery astrometric data point, along with our own astrometric measurement, to determine if it is possible to draw conclusions on the proper motion of the object relative to the primary star. The corresponding diagram is shown in Fig.~\ref{wasp76-pm}. In order to achieve an accurate position measurement of this faint source, we employed high pass filtering on the images to remove the bright halo of the exoplanet host. \\
Due to the short time baseline of only two months, and the large uncertainties given by \cite{2015AandA...579A.129W} (presumably due to worse weather conditions compared with our own detection), it is not possible to draw firm conclusions on the proper motion of the companion candidate. However, we note that our own measurement is in principle more consistent with the object being a non-moving background source rather than a bound companion. Particularly the 1$\sigma$ deviation of the two separation measurements could be well explained by parallactic displacement of the primary star relative to a presumably distant background source. Any future measurement with a similar precision as our own measurement of August 2014 will be enough to determine the status of this companion candidate.  \\
\\
\textit{HD\,185269}\\
\\
A low-mass companion to the HD\,185269 system was discovered by us with AstraLux observations in \cite{2012MNRAS.421.2498G} with observations performed between 2008 and 2011. We followed up on this companion in our current study with observations taken in July 2013 and August 2014. We show the image obtained in the 2013 observation epoch in Fig.~\ref{known-comp-images}.
In this observation epoch we observed for the first time that the companion appeared extended in north-east/south-west direction, while the PSF of the primary star showed no such distortion. This prompted us to re-observe this system with SPHERE/IRDIS. The much higher resolution extreme AO images of SPHERE show for the first time that the companion is actually a very low mass binary system itself with two approximately equally bright components (see Fig.~\ref{hd185269-sphere}).
In addition to the (unresolved) follow-up astrometry performed with AstraLux, we measured the relative position of each binary component to the primary star in all bands of the SPHERE/IRDIS observation. We used again Gaussian fitting to determine the positions of all objects.
The primary star shows a very mild saturation of the innermost 2-3 pixels in Y and H-band. We measured its position again multiple times to ensure that we reached a good fit (we fit the flanks of the saturated PSF in this case). Final results are listed in Tab.~\ref{tab: hd185269-sphere}. In addition, we used our measurements to calculate the weighted average of the position of the Bb component with respect to the Ba component. We arrive at a separation of 123.55$\pm$0.44\,mas ($\sim$5\,au projected separation at a distance of 47.37$\pm$1.72\,pc, \citealt{2007AandA...474..653V}) and a position angle of 214.87$\pm$0.21\,deg.\\
Since the SPHERE image confirmed that HD\,185269\,B is a binary, we re-examined our 2013 AstraLux observation in order to provide an astrometric measurement of the relative binary position. 
This is useful to determine the orbit of the binary and constrain its mass dynamically in later follow-up studies of the system. Due to the marginally resolved nature of the binary source in our 2013 AstraLux data, Gaussian fitting proved to be difficult. 
Instead we used the primary star's PSF as template and fitted it to the two components of HD185269\,B using IDL starfinder. This fit yielded a separation of 95.6$\pm$2.8\,mas and a PA of 221.1$\pm$1.3\,deg of Bb relative to Ba, as well as separations of 4538$\pm$14\,mas and 4458$\pm$14\,mas and PAs of 8.39$\pm$0.17\,deg and 7.72$\pm$0.18\,deg of Ba and Bb relative to A.
As expected for a system with such small separation, we see strong orbital motion between the 2013 and the 2015 observation epoch. Due to the non-optimal weather conditions in 2014, the companion is not resolved in our 2014 AstraLux observation. 
At least one additional astrometric measurement is needed to constrain the orbital elements of this binary system.\\
\\
\textit{HD\,43691}\\
\\
HD\,43691 was imaged by us once in March of 2015. We detected a companion candidate approximately 4.4\,arcsec to the north-east of the exoplanet host star. 
Since we only have one epoch it is not yet possible to determine if the object is indeed related to the HD\,43691 system. However, upon close inspection of the companion candidate's PSF we noticed that it appears extended along an angle of roughly 135\,deg. 
A close-up of the companion candidate's PSF, as well as the primary stars' PSF, is shown in Fig.~\ref{hd43691-contour}. We actually see at least two distinct peaks in the PSF (signal-to-noise ratio\footnote{The noise was determined by calculating the standard deviation in a 5$\times$5 pixel box centered on the two brightest peaks of the source.} of 5.8 and 5.5, separation of $\sim$84\,mas, i.e. 6.7\,au at 80.4\,pc), which would indicate that the object itself may be a multiple system. We compared the companion candidate's PSF with the PSF of the primary star to exclude that this is merely an effect caused by the observation conditions. 
However, the primary star' PSF appears circular in the center with a halo that is slightly extended in north-south direction, i.e. we see no indication for an intrinsic smearing of the PSF along the angle seen in the companion candidate. We note that there appears to be a third peak directly north of the south-east component of the companion candidate's PSF. This might indeed be a residual of a north-south extended halo, as seen in the primarie's PSF.
The object might hence be a binary or even trinary companion to HD\,43691\,A. However, further observations are required to confirm that the source is co-moving with the primary star and that it is indeed a multiple system itself.\\
\\
\textit{Kepler-37}\\
\\
Kepler-37 (KOI-245, KIC 8478994) was observed by us only once in August of 2014. In this data set we discovered a wide ($\sim$8.5\,arcsec) companion candidate south-south-west of the exoplanet host star. Kepler-37 was previously observed by \cite{2014A&A...566A.103L}, also using AstraLux at the Calar Alto observatory. In addition, it was targeted by \cite{2012AJ....144...42A} using ARIES at the MMT observatory. Both studies do not mention the companion candidate recovered in our own AstraLux image, since they are focusing on close companions within 6\,arcsec of the primary star.\\
Since the object was located at such a relatively large separation, we decided to check the 2MASS (\citealt{2006AJ....131.1163S}) survey for previous detection. 
While the object was not listed in the 2MASS point source catalog, it was visible in the reduced J, H and K images. We extracted the astrometric position from the individual 2MASS images using Richardson-Lucy deconvolution and then averaged the results over all bands. For details on the extraction we refer to our recent study \cite{2015MNRAS.450.3127M}. 
We find a separation of 8.030$\pm$0.138\,arcsec and a PA of 202.99$\pm$1.85\,deg in the 2MASS observation epoch of 1998.47. We used the 2MASS data in combination with our more precise AstraLux measurement to test if the discovered object is co-moving with the primary star. 
The corresponding diagram is shown in Fig.~\ref{kepler37-pm}. Even though the uncertainties of the 2MASS measurement are large compared to our AstraLux measurement, the position of the companion candidate in the 2MASS epoch is consistent within 1$\sigma$ with a non-moving background object. By comparison, co-motion with the primary can be rejected on the 4$\sigma$ level.
We thus conclude that the object is likely located in the distant background and is not physically associated with the Kepler-37 system.\\
\\
\textit{Kepler-21}\\
\\
Kepler-21 (KOI-975, KIC 3632418) was observed in July of 2013 and August 2014 with Astralux. A very close companion candidate at approximately 0.8\,arcsec was detected south-east of the primary star. In order to get an accurate position measurement of this faint source, we employed high pass filtering on the images to remove the bright halo of the exoplanet host.
The resulting measurements were compared with the proper motion of the primary star. The corresponding diagram is shown in Fig.~\ref{kepler21-pm}. Due to the direction of motion of Kepler-21, no significant change in separation would be expected for a co-moving object as well as a non-moving background object.
However, as can be seen in the diagram, both types of objects diverge in expected PA. Our measurements of the PA of the companion candidate show no significant change in PA, consistent with common proper motion. We can reject the background hypothesis with 4.0$\sigma$. 
We thus conclude that the object that we detected is most likely gravitationally bound to Kepler-21\,A and is thus a new low-mass stellar companion in this system.\\
\\
\textit{Kepler-68}\\
\\
Kepler-68 (KOI-246, KIC 11295426) was imaged by us also in July of 2013 and August of 2014. A wide companion candidate approximately 11\,arcsec to the south-east was detected in both observation epochs. 
Unfortunately Kepler-68 exhibits only a very small proper motion of -10.60$\pm$1.60 mas/yr in declination and -8.50$\pm$1.60 mas/yr in right ascension. Thus with only one year of epoch difference it was not possible to assert whether the companion candidate is co-moving with the primary star. 
However, since the companion candidate is located at a wide separation, we checked again the 2MASS catalogue to see if the source had been previously detected. 
We found that our companion candidate is indeed contained in the 2MASS point source catalogue at a relative position of 10.989$\pm$0.085\,arcsec and 145.45$\pm$0.44\,deg. Using this additional observation epoch we tested the companion for common proper motion with the primary star. 
The corresponding diagram is shown in Fig~\ref{kepler68-pm}. While the separation is inconclusive due to the tangential direction of motion of the primary star, we would have expected a significant change in PA of our measurements with respect to the 2MASS epoch if the companion candidate was a non-moving background object.
Instead we find that all measurements are consistent with no change in PA. However, due to the large uncertainties of the 2MASS epoch we can only reject the background hypothesis with 2.1$\sigma$. We conclude that, given our data, it seems likely that the companion candidate is indeed bound to the Kepler-68 system, but further observations need to be undertaken to strengthen this conclusion.\\
\\
\textit{HD\,188015}\\
\\
HD\,188015 was observed by us with AstraLux in July 2013 and August 2014. In Fig.~\ref{hd188015} we show our 2013 observation epoch. A total of 6 companion candidates are visible in the field of view of AstraLux. The high density of objects in the field of view compared to other systems is not entirely surprising since HD\,188015 is located in the direction of the Galactic disk (Galactic latitude of +00.5428$^\circ$). We note that HD\,188015 has a known low-mass stellar companion at $\sim$13\,arcsec and a PA of 85\,deg, discovered by \cite{2006ApJ...646..523R} in Sloan Digital Sky Survey (SDSS, \citealt{2000AJ....120.1579Y}) data. This companion is outside of the field of view of our AstraLux observations.\\          
In August 2014 observation conditions were not as favorable as in 2013 with shorter coherence times and thin cloud layers passing through during the observations. Thus only the candidate marked as cc1 was re-detected with high signal-to-noise. Of the other five companions, four were detected marginally with cc5 being the exception. The marginal detections in 2014 did not allow for fitting of a Gaussian to the companion candidates. Instead we have determined the center of light with a simple centroid for those four sources.
This led to much larger uncertainties of the 2014 astrometry. We nonetheless used the 2014 astrometry in combination with the known proper motion of the primary star to determine if one or several of the companion candidates are co-moving with the primary star.
The corresponding diagrams for cc1 to cc6 (with the exception of cc5) are shown in Fig.~\ref{hd188015-cc1-pm} to Fig.~\ref{hd188015-cc6-pm}. For cc1 the available astrometry is more consistent with a background object and we can reject common proper motion with HD\,188015\,A on the 3$\sigma$ level. 
For the remaining companion candidates we cannot reject common proper motion or background hypothesis with any significance. This is caused by the larger uncertainty of the 2014 measurements in combination with the short time baseline of only one year. We note, however, that the astrometry of cc2 and cc4 is more consistent with a distant background object, while the same is not true for cc3 and cc6. The latter two remain completely inconclusive due to their opposite behavior in separation and PA.
We point out that this seemingly mixed behavior could be caused by a non-zero proper motion of these objects. To gain a better understanding of this system, at least one further observation epoch in good observing conditions is required.\\
\\
\textit{HD\,197037}\\
\\
HD\,19037 was observed by us with AstraLux in two epochs in July 2013 and August 2014. We detected a companion candidate approximately 3.7\,arcsec to the south of the primary star. 
Using the proper motion of the primary star, we calculated the expected position of a non-moving background object in 2013 given the 2014 measurement. 
The corresponding diagram is shown in Fig.~\ref{hd197037-pm}. The astrometry in both epochs is consistent with no significant change in relative position. We can reject the background hypothesis with 4.8$\sigma$ in separation and 18.6$\sigma$ in PA.
We conclude that the detected object is co-moving with HD\,19037\,A and is thus most likely a new gravitationally bound stellar companion to the system. \\
\\
\textit{HD\,217786}\\
\\
HD\,217786 was observed by us on three different occasions in July 2011 and 2013, as well as in August of 2014. In all three observation epochs we detected a companion candidate approximately 2.8\,arcsec to the south of the primary star. The proper motion for this system is well determined to be -88.78$\pm$0.84 mas/yr in declination and -170.13$\pm$0.61 mas/yr in right ascension (\citealt{2007AandA...474..653V}).
We show the astrometric measurements as well as the expected behavior of a background object in Fig.~\ref{hd217786-pm}. The PA of the companion candidate is not changing significantly with time. However, we detect a small increase in separation. The dashed lines in the diagram show the expected change for a circular edge-on orbit.
The data points are consistent with such a change within 1$\sigma$. We note that even stronger changes in separation are possible for eccentric orbits. The change in separation is much smaller than what would be expected from a background object and is also showing the wrong direction (for a background object the separation should have decreased from 2011 to 2014). In fact we can reject the background hypothesis with 42.8$\sigma$ in separation and 18.9$\sigma$ in PA.
We thus conclude that the discovered object is very likely bound to the system and emerges as new low-mass stellar companion. Due to the small change in separation, but no change in PA, we expect the companion to be in a close to edge-on orbit configuration, but longer astrometric monitoring is required to test this hypothesis.

\begin{figure*}
\subfloat[WASP-76]{
\includegraphics[scale=0.38]{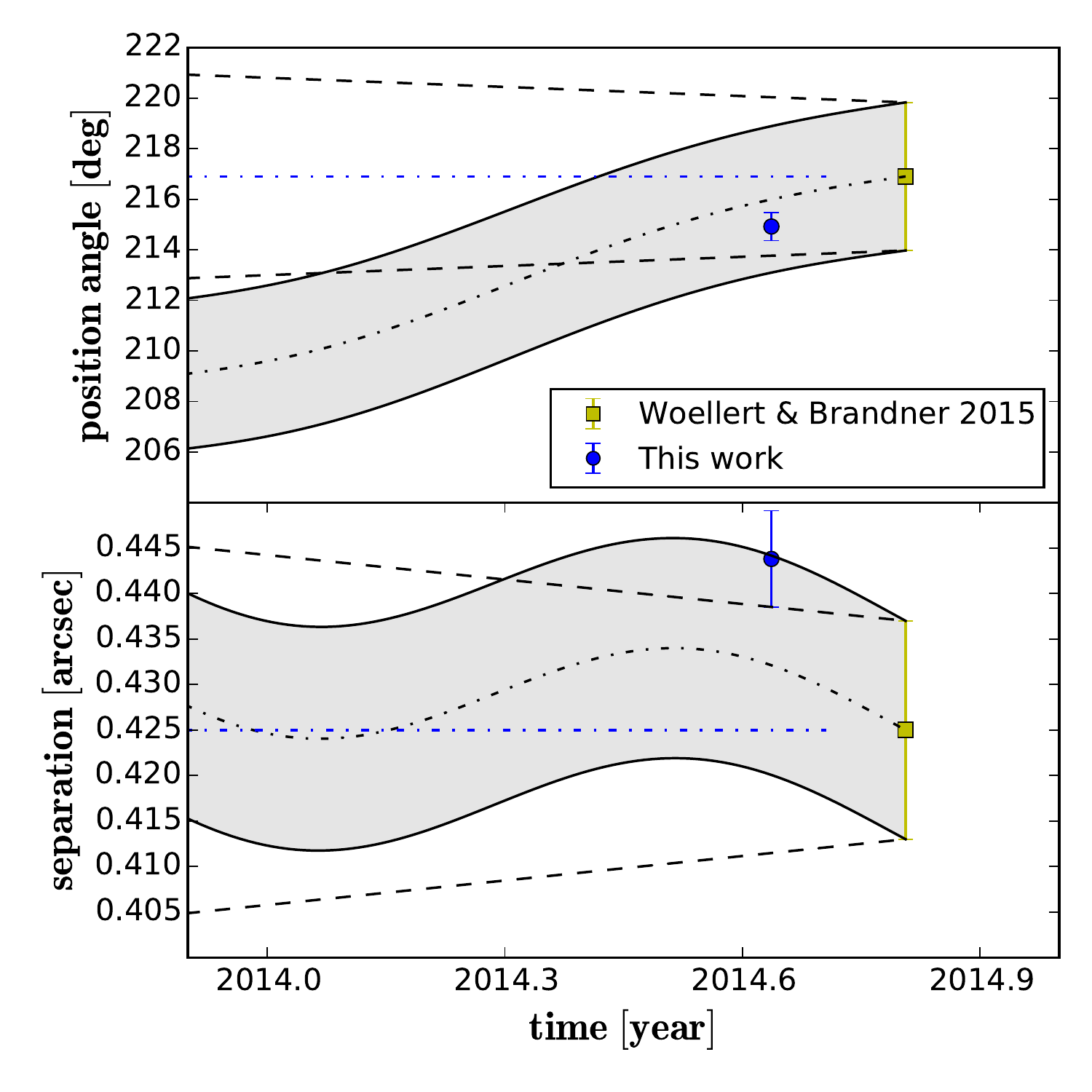}
\label{wasp76-pm}
}
\subfloat[Kepler-37]{
\includegraphics[scale=0.38]{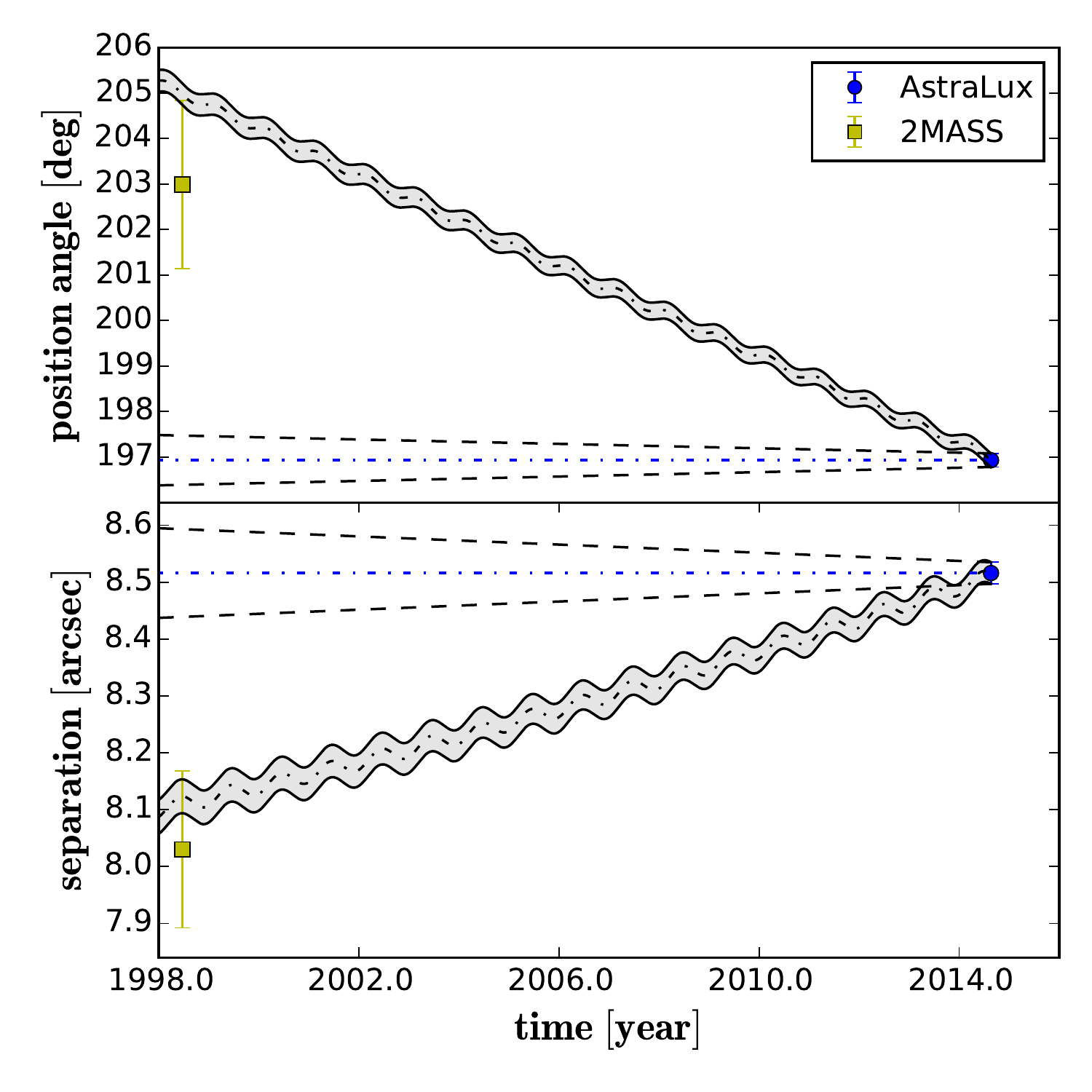}
\label{kepler37-pm}
}
\subfloat[Kepler-21]{
\includegraphics[scale=0.38]{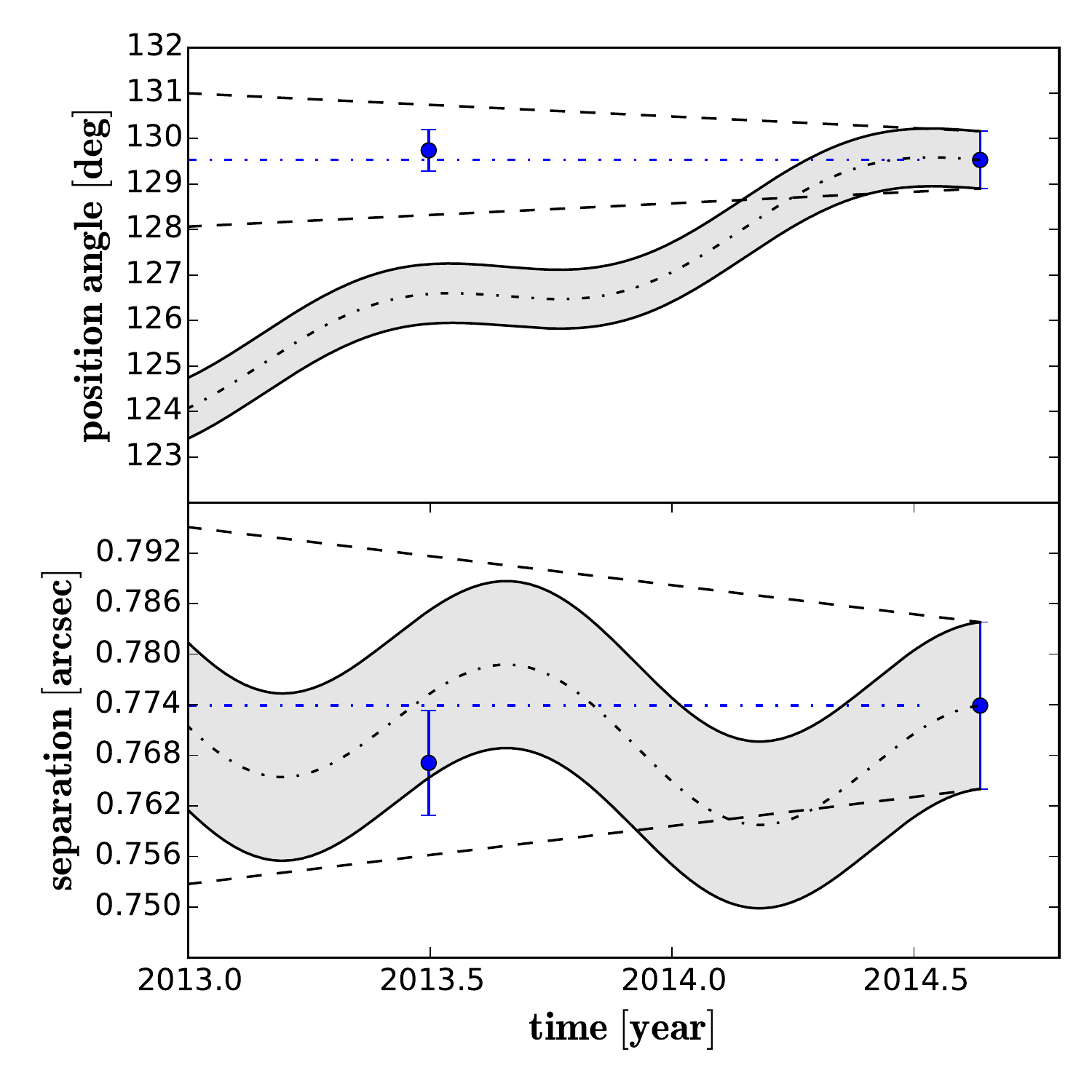}
\label{kepler21-pm}
}

\subfloat[Kepler-68]{
\includegraphics[scale=0.38]{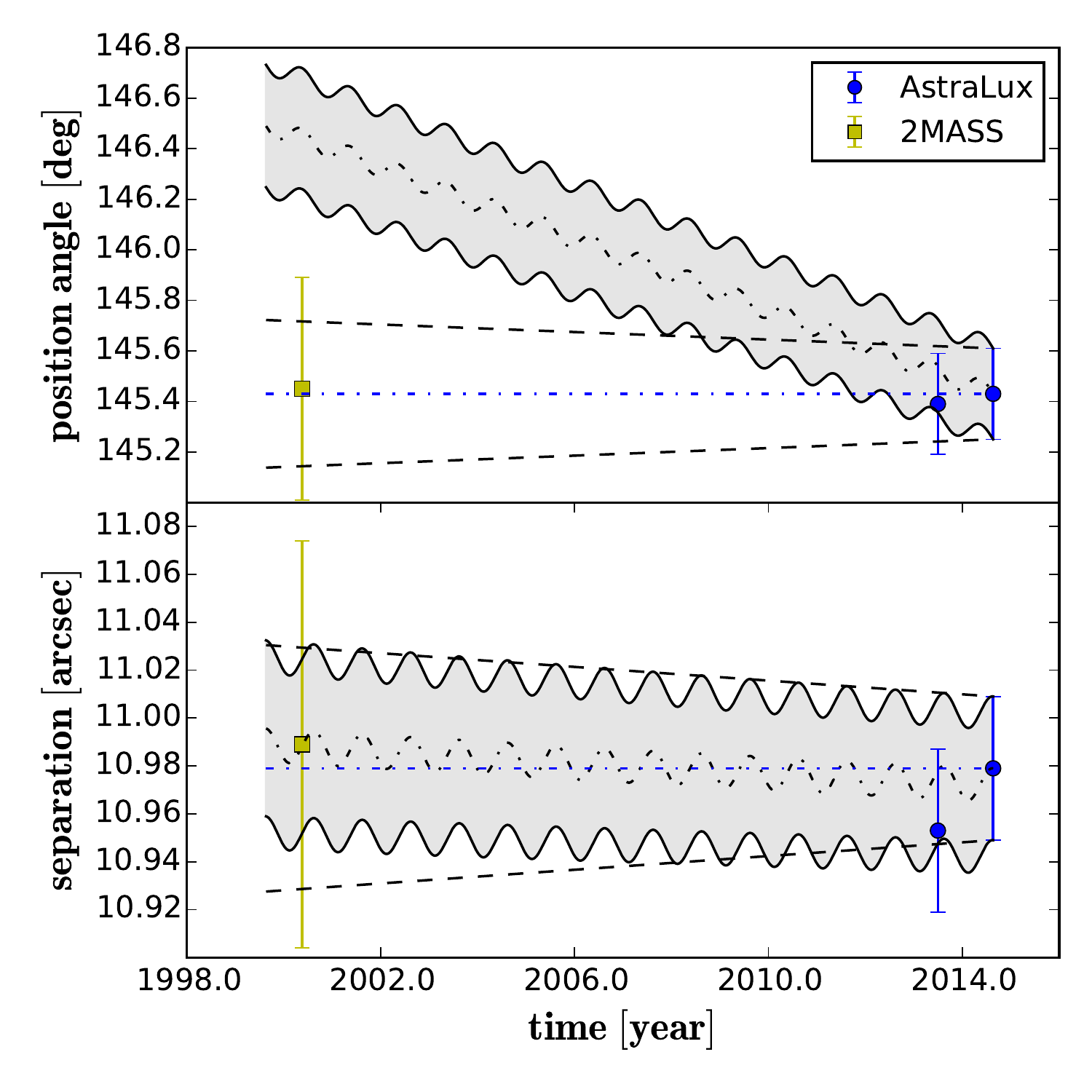}
\label{kepler68-pm}
}
\subfloat[HD\,188015\,cc\,1]{
\includegraphics[scale=0.38]{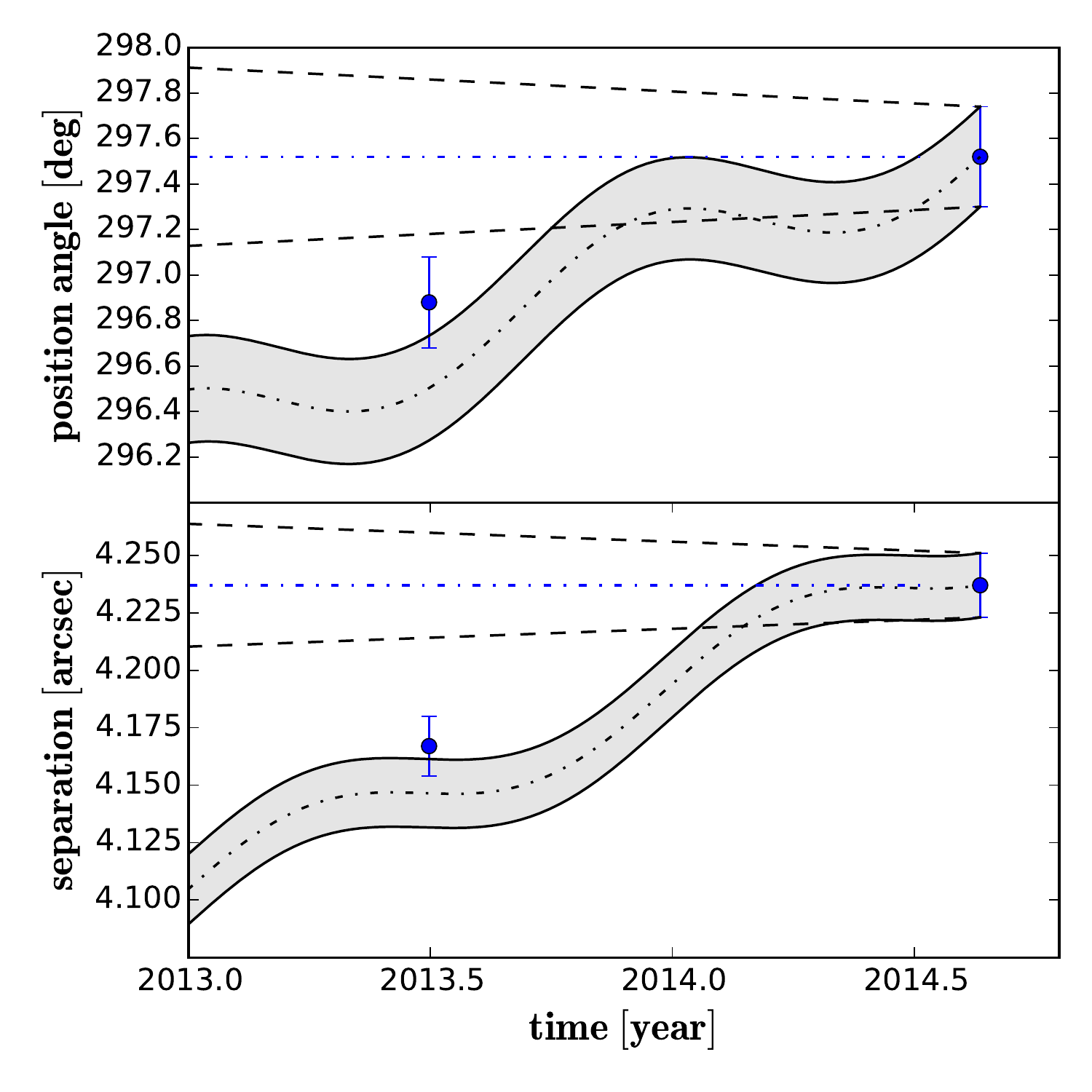}
\label{hd188015-cc1-pm}
}
\subfloat[HD\,188015\,cc\,2]{
\includegraphics[scale=0.38]{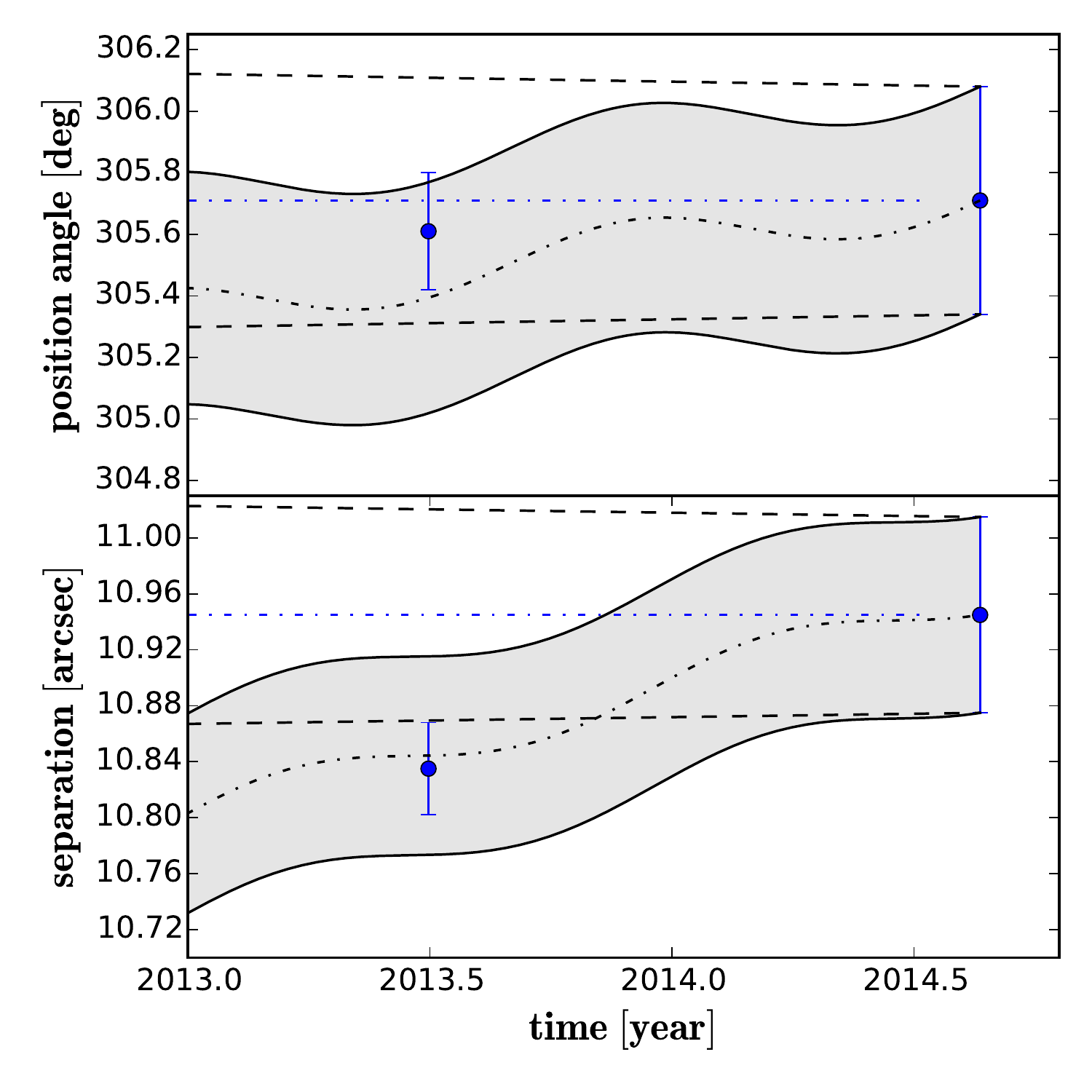}
\label{hd188015-cc2-pm}
}

\subfloat[HD\,188015\,cc\,3]{
\includegraphics[scale=0.38]{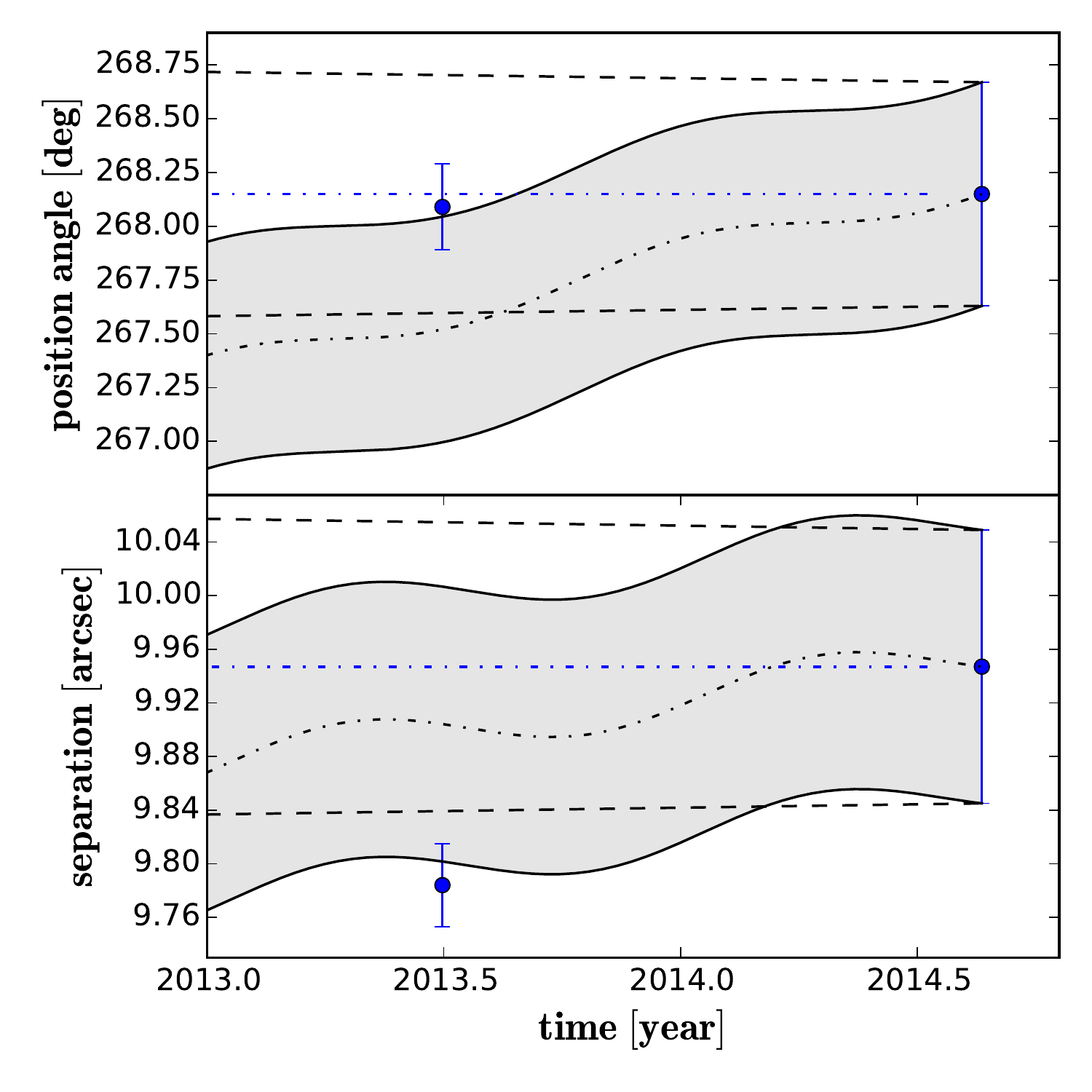}
\label{hd188015-cc3-pm}
}
\subfloat[HD\,188015\,cc\,4]{
\includegraphics[scale=0.38]{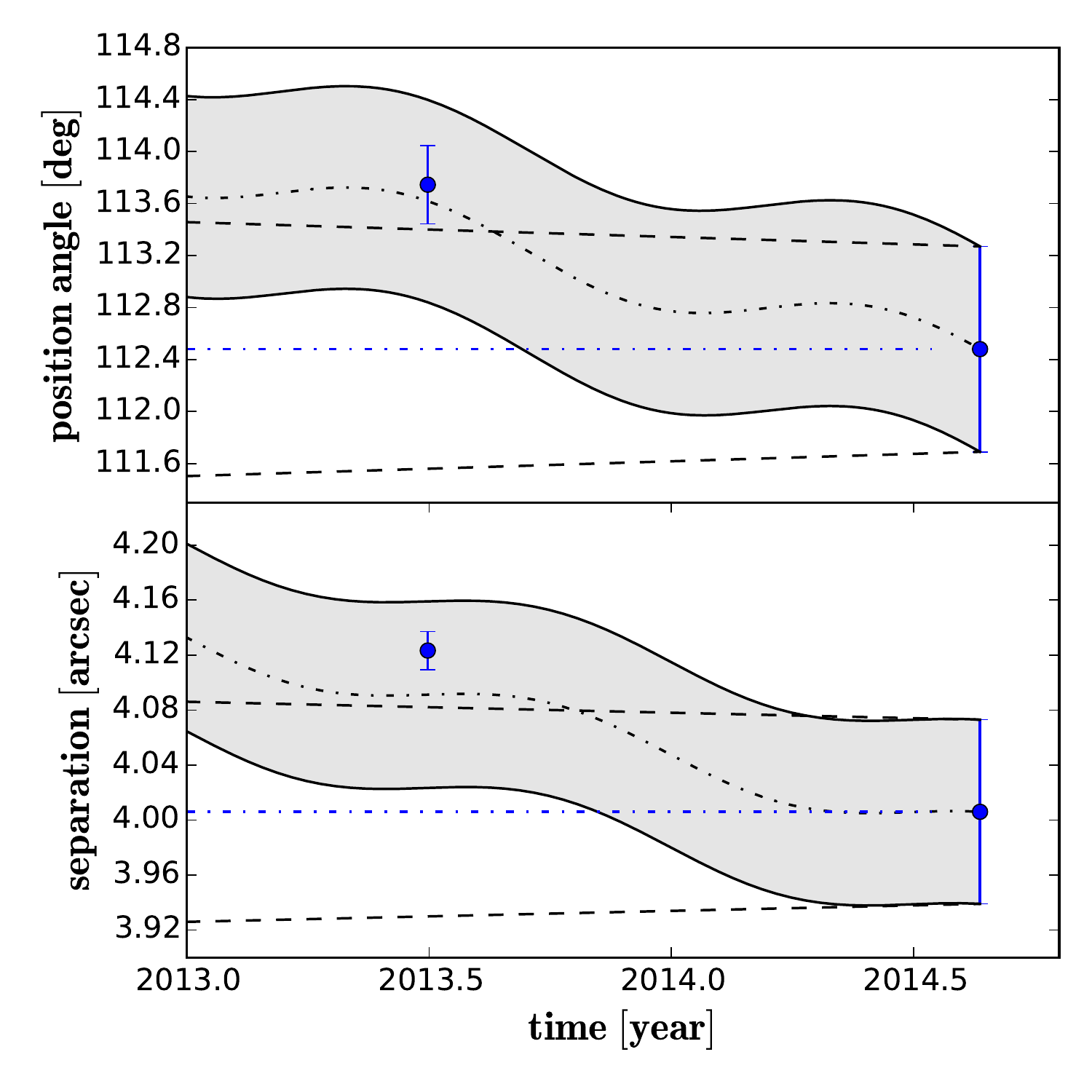}
\label{hd188015-cc4-pm}
}
\subfloat[HD\,188015\,cc\,6]{
\includegraphics[scale=0.38]{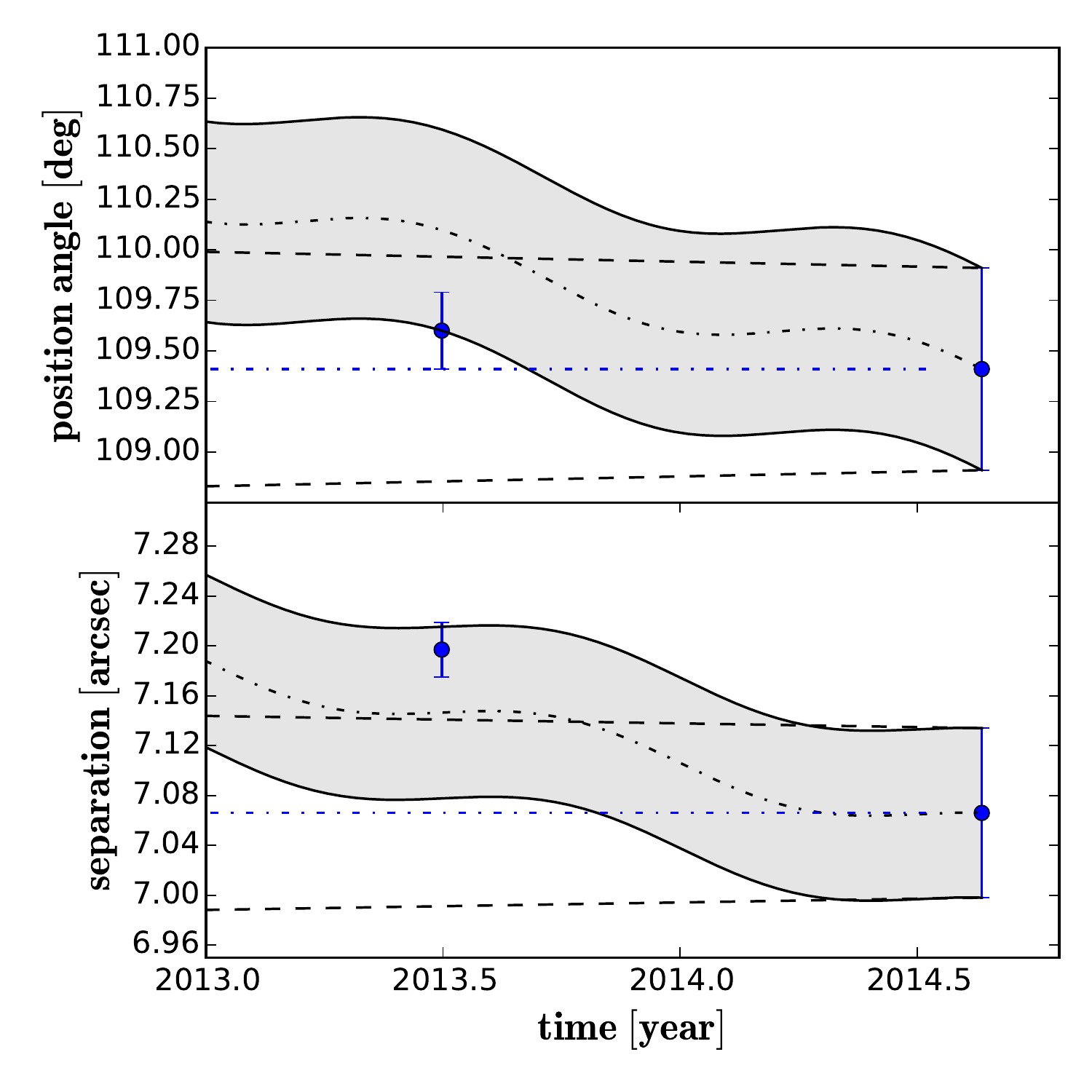}
\label{hd188015-cc6-pm}
}

%\phantomcaption
\caption[]{Proper motion analysis for all companion candidates with two or more observation epochs. Data points are AstraLux measurements if not otherwise marked. The dashed lines enclose the area in which a co-moving companion would be expected. This takes into account possible circular orbital motion with the semi-major axis given by the projected separation of the companion.
The grey area enclosed by the wobbled lines is the area in which a non-moving background object would be expected, depending on the proper motion and distance of the primary star. The wobble is introduced by the parallactic shift in the primary position due to the Earth's revolution around the sun.} 

\end{figure*}
\begin{figure*}

\ContinuedFloat

\subfloat[HD\,197037]{
\includegraphics[scale=0.38]{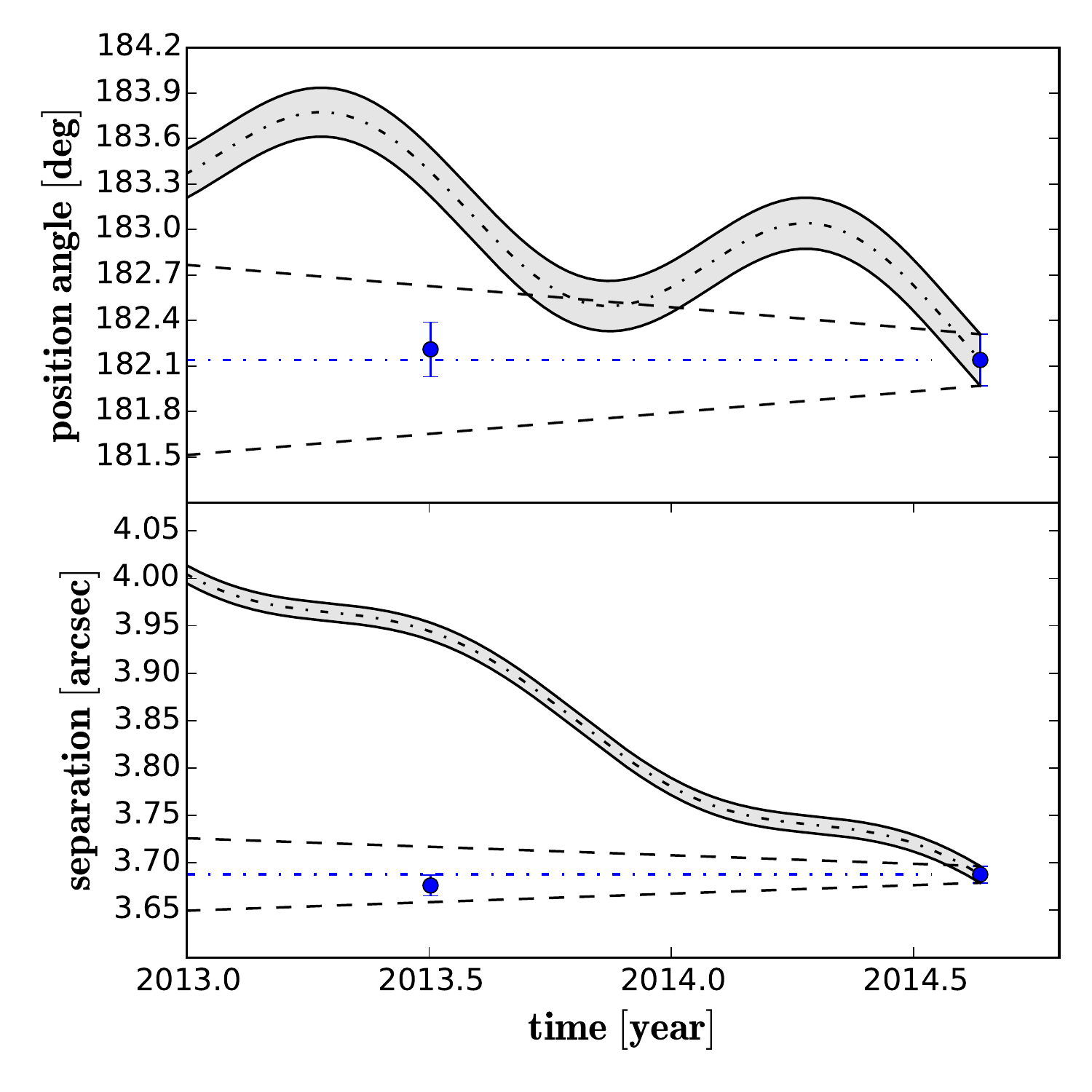}
\label{hd197037-pm}
}
\subfloat[HD\,217786]{
\includegraphics[scale=0.38]{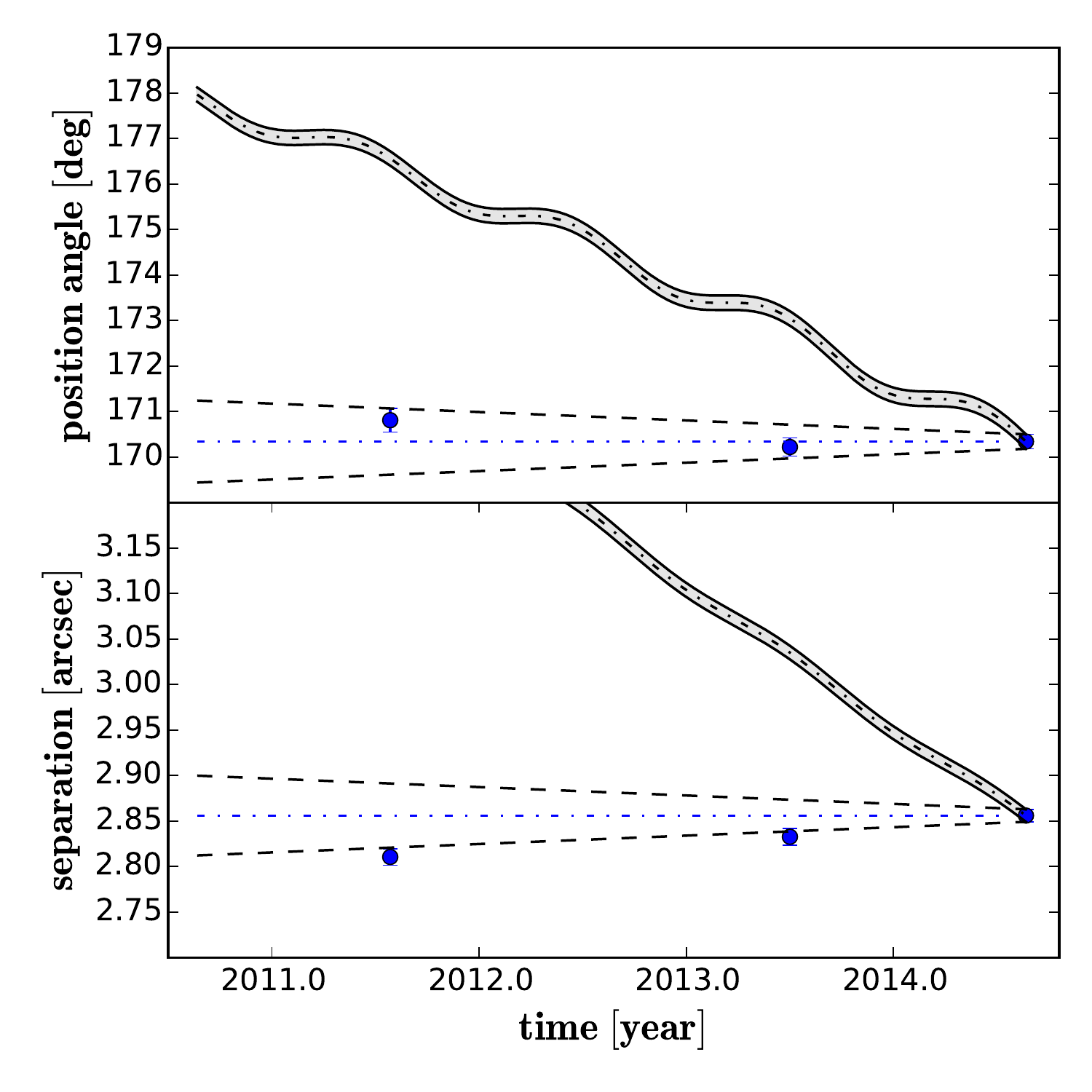}
\label{hd217786-pm}
}

\caption[]{\textbf{Continued.} Proper motion analysis for all companion candidates with two or more observation epochs. Data points are AstraLux measurements if not otherwise marked. The dashed lines enclose the area in which a co-moving companion would be expected. This takes into account possible circular orbital motion with the semi-major axis given by the projected separation of the companion.
The grey area enclosed by the wobbled lines is the area in which a non-moving background object would be expected, depending on the proper motion and distance of the primary star. The wobble is introduced by the parallactic shift in the primary position due to the Earth's revolution around the sun.} 
\label{pm-diagrams}
\end{figure*}

\begin{table}
 %\centering
 %\begin{minipage}{200mm}
 \caption{Astrometric calibration of all observation epochs as derived from observations of the center of the globular cluster M\,15. During our 2015 observation epoch M\,15 was not visible; we instead used binary stars. We list the pixel-scale (PS) and the position angle (PA) of the y-axis for all observation epochs. }
 %\begin{threeparttable}
  \begin{tabular}{@{}ccc@{}}
  \hline 
 Epoch 	& PS $[mas/pix]$&  PA of y-axis $[^\circ]$\\
 \hline
	30-06-2013		&	46.748 $\pm$ 0.14	&	358.18 $\pm$ 0.16		\\
	19-08-2014		&	46.864 $\pm$ 0.10	&	358.15 $\pm$ 0.12		\\
	10-03-2015		&	46.834 $\pm$ 0.13	&	357.66 $\pm$ 0.15		\\

\hline\end{tabular}
\label{tab: astrocal}
%\end{threeparttable}
%\end{minipage}
\end{table}

\begin{table*}
 %\centering
 %\begin{minipage}{200mm}
  \caption{Relative astrometry and photometry of all detected known companions and new companion candidates extracted from our Astralux observations. We indicate if the companion candidate is co-moving with the host star or not, if this can already be determined. We also give the confidence level of the proper motion result for the newly detected companion candidates, as well as the corresponding reference for the previously known systems.}
  \begin{tabular}{@{}lccccccc@{}}
  \hline
        
 Star 		& \# cc		& 	epoch  		& separation [arcsec] 		& 	position angle [deg] 		& 	$\Delta$mag [mag] 	& 	co-moving?	& confidence level\\
 \hline
\multicolumn{7}{c}{Known companions} \\
\hline

HD\,2638 	&		&	20-08-2014	&	0.5199 $\pm$ 0.0040	&	167.76 $\pm$ 0.35		&	3.11 $\pm$ 0.41		&	yes		&	\cite{2015AJ....149..118R}	\\
HAT-P-7		&		& 	19-08-2014	&	3.828 $\pm$ 0.011	&	89.76 $\pm$ 0.20		&	7.556 $\pm$ 0.068	&	yes		&	\cite{2010PASJ...62..779N}	\\
HD\,185269	&		&	30-06-2013	&	4.501 $\pm$ 0.016	&	8.09 $\pm$ 0.24			&	7.018 $\pm$ 0.067	&	yes		&	\cite{2012MNRAS.421.2498G}	\\
		&		&	19-08-2014	&	4.533 $\pm$ 0.014	&	8.06 $\pm$ 0.22			&	7.118 $\pm$ 0.074	&			&		\\
WASP-76		&		&	20-08-2014	&	0.4438 $\pm$ 0.0053	&	214.92 $\pm$ 0.56		&	2.58 $\pm$ 0.27		&	-		&	\cite{2015AandA...579A.129W}	\\
HAT-P-32	&		&	20-08-2014	&	2.9250 $\pm$ 0.0074	&	110.79 $\pm$ 0.17		&	5.403 $\pm$ 0.057	&	yes		&	\cite{2015ApJ...800..138N}	\\

 \hline
\multicolumn{7}{c}{New companion candidates} \\
\hline

HD\,10697	&		&	21-08-2014	&	8.858 $\pm$ 0.019	&	286.73 $\pm$ 0.14		&	7.402 $\pm$ 0.095	&	-		&		\\
HD\,43691	&		&	10-03-2015	&	4.435 $\pm$ 0.016	&	40.77 $\pm$ 0.24		&	7.71 $\pm$ 0.11		&	-		&		\\
HD\,116029	&		&	30-06-2013	&	1.3871 $\pm$ 0.0058	&	209.11 $\pm$ 0.28		&	8.8 $\pm$ 1.8		&	-		&		\\
HAT-P-18	&		&	01-07-2013	&	2.643 $\pm$ 0.014	&	185.72 $\pm$ 0.33		&	7.19 $\pm$ 0.12		&	-		&		\\
Kepler-37	&		&	20-08-2014	&	8.516 $\pm$ 0.019	&	196.93 $\pm$ 0.15		&	6.347 $\pm$ 0.056	&	no		&	4.3\,$\sigma$	\\
Kepler-21  	&		& 	02-07-2013	&	0.7671 $\pm$ 0.0062	&	129.74 $\pm$ 0.46		&	5.9$^{+4.2}_{-1.0}$	&	yes		&	4.0\,$\sigma$	\\
		&		&	20-08-2014	&	0.7739 $\pm$ 0.0099	&	129.53 $\pm$ 0.63		&	$<$ 8.1			&			&		\\
Kepler-68	&		&	02-07-2013	&	10.953 $\pm$ 0.034	&	145.39 $\pm$ 0.20		&	6.569 $\pm$ 0.073	&	yes		&	2.1\,$\sigma$	\\
		&		&	19-08-2014	&	10.979 $\pm$ 0.030	&	145.43 $\pm$ 0.18		&	6.641 $\pm$ 0.075	&			&		\\
Kepler-42	&		&	01-07-2013	&	5.206 $\pm$ 0.017	&	118.93 $\pm$ 0.21		&	4.157 $\pm$ 0.082	&	-		&		\\
HD\,188015	&	1	&	01-07-2013	&	4.167 $\pm$ 0.013	&	296.88 $\pm$ 0.20		&	8.46 $\pm$ 0.12		&	no		&	3.0\,$\sigma$	\\
		&	2	&	01-07-2013	&	10.835 $\pm$ 0.033	&	305.61 $\pm$ 0.19		&	9.00 $\pm$ 0.15		&	-		&		\\
		&	3	&	01-07-2013	&	9.784 $\pm$ 0.031	&	268.09 $\pm$ 0.20		&	9.40 $\pm$ 0.18		&	-		&		\\
		&	4	&	01-07-2013	&	4.063 $\pm$ 0.013	&	113.72 $\pm$ 0.20		&	9.05 $\pm$ 0.15		&	-		&		\\
		&	5	&	01-07-2013	&	7.037 $\pm$ 0.021	&	168.55 $\pm$ 0.19		&	9.35 $\pm$ 0.18		&	-		&		\\
		&	6	&	01-07-2013	&	7.197 $\pm$ 0.022	&	109.60 $\pm$ 0.19		&	8.78 $\pm$ 0.14		&	-		&		\\
		&	1	&	20-08-2014	&	4.237 $\pm$ 0.014	&	297.52 $\pm$ 0.22		&	8.91 $\pm$ 0.23		&			&		\\
		&	2	&	20-08-2014	&	10.9449 $\pm$ 0.070	&	305.71 $\pm$ 0.37		&	9.11 $\pm$ 0.23		&			&		\\
		&	3	&	20-08-2014	&	9.947 $\pm$ 0.102	&	268.15 $\pm$ 0.52		&	9.47 $\pm$	0.29	&			&		\\
		&	4	&	20-08-2014	&	4.006 $\pm$ 0.067	&	112.48 $\pm$ 0.79		&	9.25 $\pm$	0.36	&			&		\\
		&	6	&	20-08-2014	&	7.066 $\pm$ 0.068	&	109.41 $\pm$ 0.50		&	8.78 $\pm$	0.18	&			&		\\
HD\,197037	&		&	02-07-2013	&	3.676 $\pm$ 0.011	&	182.21 $\pm$ 0.18		&	5.124	$\pm$ 0.051	&	yes		&	19.2\,$\sigma$	\\
		&		&	20-08-2014	&	3.6876 $\pm$ 0.0088	&	182.14 $\pm$ 0.17		&	5.159	$\pm$ 0.052	&			&		\\
HD\,217786	&		&	28-07-2011	&	2.8105 $\pm$ 0.0091	&	170.81 $\pm$ 0.26		&	7.212	$\pm$ 0.078	&	yes		&	46.8\,$\sigma$	\\
		&		&	01-07-2013	&	2.8327 $\pm$ 0.0092	&	170.22 $\pm$ 0.20		&	7.171	$\pm$ 0.084	&			&		\\
		&		&	21-08-2014	&	2.8560 $\pm$ 0.0069	&	170.34 $\pm$ 0.16		&	7.160	$\pm$ 0.096	&			&		\\

\hline\end{tabular}
\label{tab:measurements}
%\end{minipage}
\end{table*}

\begin{table}
 %\centering
 %\begin{minipage}{200mm}
 \caption{Astrometric measurements of the two low-mass binary components of the HD185269 system relative to the host star from SPHERE data}
 %\begin{threeparttable}
  \begin{tabular}{@{\extracolsep{4pt}}ccccc@{}}
  \hline 
			& \multicolumn{2}{c}{Ba} 				& \multicolumn{2}{c}{Bb} \\
  \cline{2-3} \cline{4-5}			
 Filter 		& Sep. $[arcsec]$	&  PA $[^\circ]$ 		& Sep. $[arcsec]$	&  PA $[^\circ]$\\
	Y		&	4.549 $\pm$ 0.011	&	8.15 $\pm$ 0.15		&	4.442 $\pm$ 0.011	&	7.43 $\pm$ 0.14	\\
	J		&	4.547 $\pm$ 0.011	&	8.15 $\pm$ 0.14		&	4.436 $\pm$ 0.011	&	7.44 $\pm$ 0.15	\\
	H		&	4.547 $\pm$ 0.011	&	8.15 $\pm$ 0.15		&	4.436 $\pm$ 0.011	&	7.43 $\pm$ 0.15	\\

\hline\end{tabular}
\label{tab: hd185269-sphere}
%\end{threeparttable}
%\end{minipage}
\end{table}

\section{Photometric measurements and mass determination}
\label{sec: phot}

To determine the masses of the confirmed companions as well as the possible companion candidates, we performed photometric measurements for all our observation epochs. 
Since the photometry depends on the gain settings of the detector as well as the observation conditions and height of the target, we did not record a photometric standard star and rather give relative photometric measurements of the companions (and candidates) to their primary stars. 
While the PSFs of all sources in one image are similar, they are changing with observing conditions and elevation of the targets as well, thus it is not possible to build a reference PSF for photometric measurements from the data. We instead decided to perform aperture photometry on all sources. 
We used the aperture photometry tool (APT, \citealt{2012PASP..124..737L}) for these measurements. The aperture size was adjusted for each image individually to encircle the majority of the flux of the companion candidates. The same aperture size was then used to get the reference measurement from the primary star. 
In the cases where the faint sources were located within the bright halo of the primary star, care was taken to select a sky aperture close to the companion position to accurately subtract the contribution of the primary to the flux in the aperture. 
In the case of the primary star, we used sky apertures with large separations from the primary in order to not oversubstract flux due to halo contributions. All results are given in Tab.~\ref{tab:measurements}. \\
The presented uncertainties take into account statistical uncertainties, which were scaled with a factor of $\sqrt{2}$ to take into account the increased photometric uncertainty of electron multiplying CCDs. 
In addition, we consider uncertainties in the differential magnitudes from changing aperture sizes, i.e. if we increase or decrease the aperture radius by up to 2\,pixels. 
These were typically in the order of 0.04\,mag and were added in quadrature to the statistical uncertainties.\\   
To convert our photometric measurements to masses we used the BT-SETTL evolutionary models for low-mass stars, brown dwarfs and planets (\citealt{2011ASPC..448...91A}). 
These models take the absolute magnitude and the age of an object as input. To compute the absolute magnitude of our confirmed/possible companions we used the apparent magnitude of the host star in the SDSS i-band, as well as the distance of the host star. We then assume that the companions are of the same age as the host star.
We summarize these input values for all targets in our survey in Tab.~\ref{tab:detection-input}. To get a finer model-grid we interpolated (linearly) between different model ages and star magnitudes. The final masses for all confirmed or possible companion candidates are listed along with their derived absolute magnitude in Tab.~\ref{tab:masses}.
The listed uncertainties for the absolute magnitude include the uncertainty of the apparent magnitude of the host star, as well as the uncertainty in the measured differential magnitude and the uncertainty in the distance of the system. The uncertainties listed for the masses of the objects also account for the uncertainty of the system age.
In the following we compare our photometric measurements and mass determination for a few systems with available literature values.\\ 
\\
\textit{HD\,2638}\\
\\
For the close stellar companion to HD\,2638 we find a differential magnitude of 3.11 $\pm$ 0.41\,mag in the SDSS i-band. Using this measurement along with the age, distance and apparent magnitude of the primary star, we find a mass of 0.425$^{+0.067}_{-0.095}$ M$_\odot$ for the companion.
The companion was originally discovered by \cite{2015ApJ...799....4R} using Robo-AO in the optical. They have two measurements in the SDSS i-band and find differential magnitudes between primary and companion of 3.39\,mag and 3.19\,mag (\citealt{2015ApJ...799....4R}, \citealt{2015AJ....149..118R}). They do not provide uncertainties for these measurements. However, given our own uncertainties, both values are within 1\,$\sigma$ of our own measurement.
To compare our mass result with independent measurements, we used the K$_s$ and J-band photometric measurements of the companion, provided in the characterization paper of the object by \cite{2015AJ....149..118R}. To calculate a mass range we use again BT-SETTL models. We find an approximate mass range of 0.53\,M$_\odot$ to 0.45\,M$_\odot$. 
While this is slightly larger than our own SDSS i-band result, both measurements are consistent within our 1$\sigma$ uncertainties. The small discrepancy might be explained by a potential oversubtraction of background flux in our SDSS i-band images.\\
\\
\textit{HAT-P-7}\\
\\
We measure a differential SDSS i-band magnitude between HAT-P-7\,A and B of 7.556 $\pm$ 0.068\,mag. This value is in excellent agreement with the measurement very recently reported in \cite{2015A&A...575A..23W}, who use the same instrument setup and find a value of 7.58 $\pm$ 0.17\,mag. 
Using our differential SDSS i-band magnitude and the system parameters listed in Tab.~\ref{tab:detection-input}, we arrive at a mass of 0.205$^{+0.026}_{-0.021}$ M$_\odot$.
This mass estimate is consistent with the mass range given in the discovery paper by \cite{2010PASJ...62..779N}, who find 0.17 - 0.20\,M$_\odot$ from near infrared and optical photometry. It also agrees with the more recent mass estimated by \cite{2015ApJ...800..138N}, who find a range of  0.196 - 0.232\,M$_\odot$, also from near infrared photometry.\\
\\
\textit{HD\,185269}\\
\\
The photometric measurements in SDSS i-band of the 2013 and 2014 AstraLux observation of this system are consistent within 1$\sigma$ with the previous value published by us in \cite{2012MNRAS.421.2498G}.
Besides the unresolved SDSS i-band photometry, the SPHERE data enabled us to take photometric measurements of the individual components of HD\,185269\,B. Since we do not have additional sources in the field of view other than the primary and the binary companion, we again used aperture photometry to derive the brightness of the binary components. 
As mentioned previously the primary star is saturated in Y and H-band, thus in these bands we could only measure the brightness difference between the binary components. However, our J-band data is unsaturated, which enabled photometric calibration of the binary measurements with the primary star.
We list all our results in Tab.~\ref{tab:hd185269-sphere-phot}. The given J-band magnitudes are assuming that the neutral density filter is flat across the covered wavelength range.\\ 
We used again the BT-SETTL models to convert the J-band measurements into masses of the individual components. For this conversion we utilized the J-band magnitude of the primary of 5.518$\pm$0.027\,mag (\citealt{2003yCat.2246....0C}) and the most recent age estimate by \cite{2013A&A...551L...8P} of 3.49$\pm$0.79\,Gyr. The system is located at a distance of 47.37$\pm$1.72\,pc (\citealt{2007AandA...474..653V}).
Given the brightness of the binary components, we calculate masses of 0.165$\pm$0.008\,M$_\odot$ and 0.154$^{+0.009}_{-0.008}$\,M$_\odot$ for the Ba and Bb components respectively. \\
To compare these results with our unresolved SDSS i-band measurements we calculated the flux ratio between the two components in Y-band. Given the differential brightness measured in our SPHERE image, the flux ratio between the two components is 0.8. 
To verify that this flux ratio is consistent with our AstraLux observations in SDSS i-band, we used the results obtained from PSF fitting of the two components of HD185269\,B in the 2013 AstraLux data mentioned in section \ref{sec:astrometry}. This PSF fitting yields a flux ratio of 0.78, i.e. consistent with the Y-band results obtained with SPHERE.
We calculated the expected apparent SDSS i-band magnitudes for both components to be 14.12$\pm$0.10\,mag and 14.36$\pm$0.10\,mag. 
From these measurements we derived SDSS i-band masses of 0.18$\pm$0.01\,M$_\odot$ and 0.16$\pm$0.01\,M$_\odot$ for the Ba and Bb component respectively. These are consistent with our more precise J-band masses within 1$\sigma$.\\
\\
\textit{WASP-76}\\
\\
For WASP-76 we find a differential magnitude in the SDSS i-band of 2.58 $\pm$ 0.27\,mag. This is very consistent with the value of 2.51 $\pm$ 0.25\,mag recovered by \cite{2015AandA...579A.129W}, using the same instrument setup. 
Given our differential magnitude and the age, distance and apparent magnitude of the primary star shown in Tab.~\ref{tab:detection-input}, we compute a mass of the object of 0.692$^{+0.074}_{-0.059}$ M$_\odot$, assuming that it is indeed gravitationally bound.\\
\\
\textit{HAT-P-32}\\
\\
In the case of the HAT-P-32 system we find a differential SDSS i-band magnitude of 5.403 $\pm$ 0.057\,mag for the stellar companion. Given the age, distance and apparent magnitude of the primary star this translates into a mass of 0.340$^{+0.048}_{-0.024}$ M$_\odot$. The mass of HAT-P-32\,B was also recently estimated by \cite{2015ApJ...800..138N}, who detected the companion in J, H and K-band. They arrived at a mass of 0.393 $\pm$ 0.012\,M$_\odot$ for J and K-band, and 0.4243 $\pm$ 0.0085\,M$_\odot$ for H-band. Our mass estimate is lower but marginally consistent within 1$\sigma$ with their J and K-band results. We are deviating from their higher H-band mass by 1.5\,$\sigma$. However, we want to point out that their two mass estimates also deviate by a similar margin. In principle it is possible that our slightly lower mass estimate is caused by an overestimation of the background, which is dominated by the bright stellar halo, even though we get consistent photometric results with other studies for sources at even smaller separations, such as WASP-76. 

\section{Detection limits}

\label{sec: detect}

\begin{figure}

\includegraphics[scale=0.42]{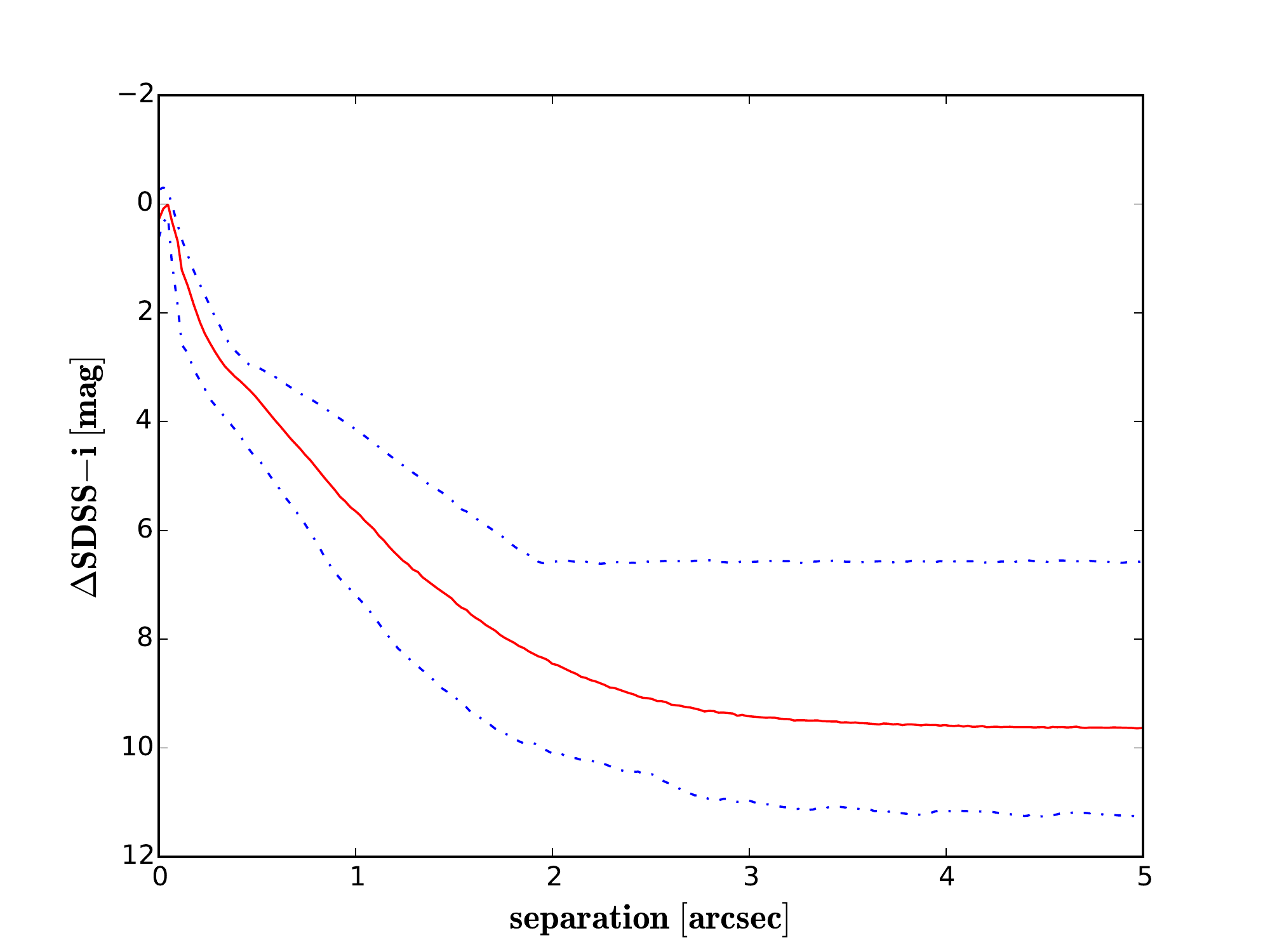}

\caption[]{Contrast achieved in our AstraLux observations. We show the average contrast (solid, red line) as well as the best and worst contrast (dash-dotted, blue lines). The contrast depends strongly on the observing conditions, which explains the large spread between the best and worst contrast. Individual contrast curves for each target are available as supplementary online material.} 
\label{average-contrast}
\end{figure}

To guide future observations and enable more sophisticated statistical analysis of the multiplicity ratio of exoplanet hosts, we have derived detection limits at various separations for each of our target stars. For this purpose, we first computed the achievable magnitude difference (contrast) compared to the bright primary star at these separations. 
We assume that an object is detectable when its' signal-to-noise ratio is equal or larger than 5. We then use the peak brightness of the bright primary star as calibration value for the signal. The noise at each separation is determined by averaging over the standard deviation measured in 5$\times$5 pixel boxes which are centered on each pixel with the respective separation from the primary star. 
In Fig.~\ref{average-contrast} we show the average contrast of all our observations along with the best and worst contrast achieved up to a separation of 5\,arcsec, at which we reach the background limit.
To convert from these magnitude limits to mass limits, we again utilized the BT-SETTL models as described in section \ref{sec: phot}. The input values for this conversion are given in Tab.~\ref{tab:detection-input}. The final derived mass limits are given in Tab.~\ref{tab:detection-output}. 
In some cases not all necessary input values were available; we then give only the achievable magnitude limit, which can be used to calculate mass limits at a later time, should all the input values become available. In addition, in a few cases the detectable minimum mass was located outside of our model grid. We then give a lower or upper detection limit based on the closest grid value.\\
Our detection limits depend mostly on the atmospheric conditions during the observations as well as the brightness and distance of the exoplanet host. Since our sample consists mostly of evolved systems with typical ages in the order of a few Gyr, the dependency of the detectable mass limit on the age is less important.
We are on average sensitive to masses down to 0.52\,$M_\odot$ outside of 1\,arcsec and down to 0.16\,$M_\odot$ in the background limited region outside of 5\,arcsec. These detection limits are comparable to our previous study \cite{2012MNRAS.421.2498G} in which we used AstraLux on a similar sample of target systems.

\begin{table*}
 %\centering
 %\begin{minipage}{200mm}
  \caption{Distances, apparent magnitudes and ages of all target stars in our survey. We give the corresponding references in adjacent columns.}
  \begin{tabular}{@{}lclclcl@{}}
  \hline
        
 Star 				& SDSS i [mag]	& Ref. 							& Distance [pc] 				& Ref. 							& Age [Gyr]			& Ref. 											\\
 \hline

HD2638	&	9.01\,$\pm$\,0.03		&	\cite{2008PASP..120.1128O}	&	49.9$\,\pm\,$4.0		&	\cite{2007AandA...474..653V}	&	1.9$\,\pm\,$2.6	&	\cite{2015AandA...575A..18B}	\\
HD2952	&		...	&		&	114.2$\,\pm\,$6.2		&	\cite{2007AandA...474..653V}	&	...	&		\\
HD5608	&	5.49\,$\pm$\,0.03		&	\cite{2008PASP..120.1128O}	&	56.4$\,\pm\,$1.3		&	\cite{2007AandA...474..653V}	&	...	&		\\
HD5891	&	7.474\,$\pm$\,0.01		&	\cite{2012ApJS..203...21A}	&	251.3$\,\pm\,$109.8		&	\cite{2007AandA...474..653V}	&	1.5$\,\pm\,$0.1	&	\cite{2015AandA...575A..18B}	\\
HD8574	&	6.97\,$\pm$\,0.01		&	\cite{2008PASP..120.1128O}	&	44.6$\,\pm\,$1.1		&	\cite{2007AandA...474..653V}	&	5.0$\,\pm\,$0.1	&	\cite{2015AandA...575A..18B}	\\
HD10697	&	5.91\,$\pm$\,0.15		&	\cite{2008PASP..120.1128O}	&	32.6$\,\pm\,$0.5		&	\cite{2007AandA...474..653V}	&	7.1$\,\pm\,$0.1	&	\cite{2015AandA...575A..18B}	\\
WASP-76	&	9.318\,$\pm$\,0.001		&	\cite{2012ApJS..203...21A}	&	120.0$\,\pm\,$20.0		&	\cite{2013arXiv1310.5607W}	&	5.3$^{+6.1}_{-2.9}$	&	\cite{2013arXiv1310.5607W}	\\
HAT-P-32	&	11.12\,$\pm$\,0.08		&	\cite{2008PASP..120.1128O}	&	320.0$\,\pm\,$16.0		&	\cite{2011ApJ...742...59H}	&	0.1$\,\pm\,$0.1	&	\cite{2015AandA...575A..18B}	\\
HD12661	&	7.1\,$\pm$\,0.04		&	\cite{2008PASP..120.1128O}	&	35.0$\,\pm\,$0.8		&	\cite{2007AandA...474..653V}	&	1.8$\,\pm\,$0.5	&	\cite{2015AandA...575A..18B}	\\
HD13189	&	6.56\,$\pm$\,0.09		&	\cite{2008PASP..120.1128O}	&	561.8$\,\pm\,$390.6		&	\cite{2007AandA...474..653V}	&	...	&		\\
HD13908	&	7.33\,$\pm$\,0.02		&	\cite{2008PASP..120.1128O}	&	71.2$\,\pm\,$3.7		&	\cite{2007AandA...474..653V}	&	2.9$\,\pm\,$0.4	&	\cite{2014AandA...563A..22M}	\\
HD15779	&	4.82\,$\pm$\,0.04		&	\cite{2008PASP..120.1128O}	&	81.4$\,\pm\,$3.1		&	\cite{2007AandA...474..653V}	&	...	&		\\
HD285507	&	9.91\,$\pm$\,0.11		&	\cite{2008PASP..120.1128O}	&	41.3$\,\pm\,$4.0		&	\cite{2007AandA...474..653V}	&	0.63$\,\pm\,$0.05	&	\cite{2014ApJ...787...27Q}	\\
HD290327	&	8.62\,$\pm$\,0.03		&	\cite{2008PASP..120.1128O}	&	56.7$\,\pm\,$5.5		&	\cite{2007AandA...474..653V}	&	11.8$\,\pm\,$1.2	&	\cite{2015AandA...575A..18B}	\\
HD40979	&	6.57\,$\pm$\,0.02		&	\cite{2008PASP..120.1128O}	&	33.1$\,\pm\,$0.5		&	\cite{2007AandA...474..653V}	&	1.5$\,\pm\,$0.5	&	\cite{2007AandA...469..755M}	\\
HD43691	&	7.88\,$\pm$\,0.02		&	\cite{2008PASP..120.1128O}	&	80.4$\,\pm\,$5.7		&	\cite{2007AandA...474..653V}	&	3.1$\,\pm\,$2.5	&	\cite{2015AandA...575A..18B}	\\
HD45350	&	7.53\,$\pm$\,0.03		&	\cite{2008PASP..120.1128O}	&	48.9$\,\pm\,$1.8		&	\cite{2007AandA...474..653V}	&	7.0$\,\pm\,$0.9	&	\cite{2015AandA...575A..18B}	\\
Omi Uma	&	2.9635\,$\pm$\,0.042		&	\cite{2005AJ....130..873J}	&	54.9$\,\pm\,$0.5		&	\cite{2007AandA...474..653V}	&	0.36$\,\pm\,$0.03	&	\cite{2008AandA...480...91S}	\\
GJ328	&	8.946\,$\pm$\,0.131		&	\cite{2005AJ....130..873J}	&	19.8$\,\pm\,$0.8		&	\cite{2013ApJ...774..147R}	&	...	&		\\
HD95089	&	7.4\,$\pm$\,0.03		&	\cite{2008PASP..120.1128O}	&	139.1$\,\pm\,$18.2		&	\cite{2007AandA...474..653V}	&	2.3$\,\pm\,$0.2	&	\cite{2015AandA...575A..18B}	\\
HD96063	&	7.76\,$\pm$\,0.04		&	\cite{2008PASP..120.1128O}	&	158.0$\,\pm\,$23.5		&	\cite{2007AandA...474..653V}	&	3.6$\,\pm\,$0.7	&	\cite{2015AandA...575A..18B}	\\
HD99706	&	7.14\,$\pm$\,0.03		&	\cite{2008PASP..120.1128O}	&	128.9$\,\pm\,$12.4		&	\cite{2007AandA...474..653V}	&	2.8$\,\pm\,$0.2	&	\cite{2015AandA...575A..18B}	\\
HD100655	&	5.93\,$\pm$\,0.11		&	\cite{2008PASP..120.1128O}	&	122.2$\,\pm\,$8.0		&	\cite{2007AandA...474..653V}	&	0.9$\,\pm\,$0.2	&	\cite{2015AandA...575A..18B}	\\
HIP57274	&	8.23\,$\pm$\,0.03		&	\cite{2008PASP..120.1128O}	&	25.9$\,\pm\,$0.7		&	\cite{2007AandA...474..653V}	&	8.4$\,\pm\,$3.7	&	\cite{2015AandA...575A..18B}	\\
HD102329	&	7.32\,$\pm$\,0.02		&	\cite{2008PASP..120.1128O}	&	158.0$\,\pm\,$23.8		&	\cite{2007AandA...474..653V}	&	2.0$\,\pm\,$0.3	&	\cite{2015AandA...575A..18B}	\\
HD106270	&	7.21\,$\pm$\,0.04		&	\cite{2008PASP..120.1128O}	&	84.9$\,\pm\,$6.1		&	\cite{2007AandA...474..653V}	&	4.0$\,\pm\,$0.1	&	\cite{2015AandA...575A..18B}	\\
HD113337	&	5.95\,$\pm$\,0.02		&	\cite{2008PASP..120.1128O}	&	36.9$\,\pm\,$0.4		&	\cite{2007AandA...474..653V}	&	0.2$\,\pm\,$0.1	&	\cite{2014AandA...561A..65B}	\\
HD116029	&	7.36\,$\pm$\,0.04		&	\cite{2008PASP..120.1128O}	&	123.2$\,\pm\,$10.7		&	\cite{2007AandA...474..653V}	&	3.5$\,\pm\,$0.5	&	\cite{2015AandA...575A..18B}	\\
HD120084	&	5.37\,$\pm$\,0.05		&	\cite{2008PASP..120.1128O}	&	100.7$\,\pm\,$2.5		&	\cite{2007AandA...474..653V}	&	1.1$\,\pm\,$0.3	&	\cite{2008AandA...480...91S}	\\
Beta UMi	&	1.081\,$\pm$\,0.042		&	\cite{2005AJ....130..873J}	&	40.1$\,\pm\,$0.2		&	\cite{2007AandA...474..653V}	&	3.0$\,\pm\,$1.0	&	\cite{2014AandA...566A..67L}	\\
HD131496	&	7.33\,$\pm$\,0.03		&	\cite{2008PASP..120.1128O}	&	110.0$\,\pm\,$10.3		&	\cite{2007AandA...474..653V}	&	4.5$\,\pm\,$0.4	&	\cite{2015AandA...575A..18B}	\\
HD131496	&	7.33\,$\pm$\,0.03		&	\cite{2008PASP..120.1128O}	&	110.0$\,\pm\,$10.3		&	\cite{2007AandA...474..653V}	&	4.5$\,\pm\,$0.4	&	\cite{2015AandA...575A..18B}	\\
HD136726	&	4.25\,$\pm$\,0.06		&	\cite{2008PASP..120.1128O}	&	122.1$\,\pm\,$2.9		&	\cite{2007AandA...474..653V}	&	3.9$\,\pm\,$0.9	&	\cite{2015AandA...575A..18B}	\\
11 UMi	&	4.25\,$\pm$\,0.06		&	\cite{2008PASP..120.1128O}	&	122.1$\,\pm\,$2.9		&	\cite{2007AandA...474..653V}	&	3.9$\,\pm\,$0.9	&	\cite{2015AandA...575A..18B}	\\
HD136512	&	4.99\,$\pm$\,0.06		&	\cite{2008PASP..120.1128O}	&	82.8$\,\pm\,$3.1		&	\cite{2007AandA...474..653V}	&	0.85$\,\pm\,$0.38	&	\cite{2008PASJ...60..781T}	\\
Omi CrB	&	4.99\,$\pm$\,0.06		&	\cite{2008PASP..120.1128O}	&	82.8$\,\pm\,$3.1		&	\cite{2007AandA...474..653V}	&	5.6$\,\pm\,$2.2	&	\cite{2008AandA...480...91S}	\\
HD139357	&	4.68\,$\pm$\,0.5		&	\cite{2003AJ....125..984M}	&	118.1$\,\pm\,$4.3		&	\cite{2007AandA...474..653V}	&	7.0$\,\pm\,$2.0	&	\cite{2015AandA...575A..18B}	\\
HD145457	&	6.08\,$\pm$\,0.11		&	\cite{2008PASP..120.1128O}	&	125.3$\,\pm\,$7.5		&	\cite{2007AandA...474..653V}	&	2.6$\,\pm\,$0.4	&	\cite{2015AandA...575A..18B}	\\
HD152581	&	7.95\,$\pm$\,0.03		&	\cite{2008PASP..120.1128O}	&	185.5$\,\pm\,$40.2		&	\cite{2007AandA...474..653V}	&	8.6$\,\pm\,$2.1	&	\cite{2015AandA...575A..18B}	\\
HAT-P-18	&	12.125\,$\pm$\,0.01		&	\cite{2009ApJS..182..543A}	&	166.0$\,\pm\,$9.0		&	\cite{2011ApJ...726...52H}	&	12.4$\,\pm\,$6.4	&	\cite{2011ApJ...726...52H}	\\
HD156279	&	7.65\,$\pm$\,0.03		&	\cite{2008PASP..120.1128O}	&	36.6$\,\pm\,$0.6		&	\cite{2007AandA...474..653V}	&	7.4$\,\pm\,$1.9	&	\cite{2015AandA...575A..18B}	\\
HD163607	&	7.643\,$\pm$\,0.001		&	\cite{2012ApJS..203...21A}	&	68.8$\,\pm\,$2.3		&	\cite{2007AandA...474..653V}	&	8.91$\,\pm\,$0.01	&	\cite{2015AandA...575A..18B}	\\
HD163917	&	2.78\,$\pm$\,0.03		&	\cite{2008PASP..120.1128O}	&	46.2$\,\pm\,$0.6		&	\cite{2007AandA...474..653V}	&	0.45$\,\pm\,$0.07	&	\cite{2008AandA...480...91S}	\\
HIP91258	&	8.33\,$\pm$\,0.02		&	\cite{2008PASP..120.1128O}	&	44.9$\,\pm\,$1.4		&	\cite{2007AandA...474..653V}	&	2.4$\,\pm\,$2.4	&	\cite{2014AandA...563A..22M}	\\
Kepler-37	&	9.38\,$\pm$\,0.04		&	\cite{2008PASP..120.1128O}	&	66.0$\,\pm\,$33.0		&	\cite{2013Natur.494..452B}	&	3.7$\,\pm\,$0.8	&	\cite{2013MNRAS.436.1883W}	\\
Kepler-21	&	8.06\,$\pm$\,0.03		&	\cite{2008PASP..120.1128O}	&	112.9$\,\pm\,$7.9		&	\cite{2007AandA...474..653V}	&	3.55$\,\pm\,$0.03	&	\cite{2015AandA...575A..18B}	\\
HD180314	&	6.14\,$\pm$\,0.05		&	\cite{2008PASP..120.1128O}	&	131.4$\,\pm\,$7.1		&	\cite{2007AandA...474..653V}	&	0.9$\,\pm\,$0.6	&	\cite{2015AandA...575A..18B}	\\
Kepler-63	&	11.44\,$\pm$\,0.02		&	\cite{2012yCat.1322....0Z}	&	200.0$\,\pm\,$15.0		&	\cite{2013ApJ...775...54S}	&	0.210$\,\pm\,$0.045	&	\cite{2013ApJ...775...54S}	\\
Kepler-68	&	9.83\,$\pm$\,0.02		&	\cite{2008PASP..120.1128O}	&	135.0$\,\pm\,$10.0		&	\cite{2013ApJ...766...40G}	&	6.3$\,\pm\,$1.7	&	\cite{2013ApJ...766...40G}	\\
Kepler-42	&	14.375\,$\pm$\,0.5		&	\cite{2012yCat.1322....0Z}	&	38.7$\,\pm\,$6.3		&	\cite{2012ApJ...747..144M}	&	5.0$\,\pm\,$1.0	&	\cite{2012ApJ...747..144M}	\\
HAT-P-7	&	10.37\,$\pm$\,0.01		&	\cite{2008PASP..120.1128O}	&	320.0$\,\pm\,$40.0		&	\cite{2008ApJ...680.1450P}	&	1.5$\,\pm\,$0.2	&	\cite{2015AandA...575A..18B}	\\
HD188015	&	7.93\,$\pm$\,0.04		&	\cite{2008PASP..120.1128O}	&	57.0$\,\pm\,$2.9		&	\cite{2007AandA...474..653V}	&	5.3$^{+2.6}_{-0.3}$	&	\cite{2012ApJ...756...46R}	\\
HD190360	&	5.41\,$\pm$\,0.04		&	\cite{2008PASP..120.1128O}	&	15.9$\,\pm\,$0.1		&	\cite{2007AandA...474..653V}	&	11.5$^{+1.3}_{-2.8}$	&	\cite{2012ApJ...756...46R}	\\
HD197037	&	6.63\,$\pm$\,0.03		&	\cite{2008PASP..120.1128O}	&	32.3$\,\pm\,$0.4		&	\cite{2007AandA...474..653V}	&	0.3$\,\pm\,$0.3	&	\cite{2015AandA...575A..18B}	\\
HD206610	&	7.87\,$\pm$\,0.04		&	\cite{2008PASP..120.1128O}	&	193.8$\,\pm\,$43.7		&	\cite{2007AandA...474..653V}	&	2.1$\,\pm\,$0.3	&	\cite{2015AandA...575A..18B}	\\
HD208527	&	4.78\,$\pm$\,0.13		&	\cite{2008PASP..120.1128O}	&	403.2$\,\pm\,$73.0		&	\cite{2007AandA...474..653V}	&	2.0$\,\pm\,$1.3	&	\cite{2013AandA...549A...2L}	\\
HD210277	&	6.23\,$\pm$\,0.04		&	\cite{2008PASP..120.1128O}	&	21.6$\,\pm\,$0.2		&	\cite{2007AandA...474..653V}	&	7.9$\,\pm\,$2.0	&	\cite{2015AandA...575A..18B}	\\
HD217786	&	7.54\,$\pm$\,0.02		&	\cite{2008PASP..120.1128O}	&	54.9$\,\pm\,$2.3		&	\cite{2007AandA...474..653V}	&	6.5$\,\pm\,$0.8	&	\cite{2015AandA...575A..18B}	\\
HD240210	&	7.16\,$\pm$\,0.13		&	\cite{2005AJ....130..873J}	&	143.0$\,\pm\,$53.0		&	\cite{2009ApJ...707..768N}	&	10.9$\,\pm\,$1.8	&	\cite{2015AandA...575A..18B}	\\
HD219828	&	7.78\,$\pm$\,0.03		&	\cite{2008PASP..120.1128O}	&	72.3$\,\pm\,$4.1		&	\cite{2007AandA...474..653V}	&	5.0$\,\pm\,$0.7	&	\cite{2015AandA...575A..18B}	\\
HD220074	&	4.77\,$\pm$\,0.07		&	\cite{2008PASP..120.1128O}	&	324.7$\,\pm\,$52.7		&	\cite{2007AandA...474..653V}	&	4.5$\,\pm\,$2.8	&	\cite{2013AandA...549A...2L}	\\
HD222155	&	6.86\,$\pm$\,0.02		&	\cite{2008PASP..120.1128O}	&	49.1$\,\pm\,$1.5		&	\cite{2007AandA...474..653V}	&	7.9$\,\pm\,$0.1	&	\cite{2015AandA...575A..18B}	\\
		
\hline\end{tabular}
\label{tab:detection-input}
%\end{minipage}
\end{table*}

\begin{table}
 %\centering
 %\begin{minipage}{200mm}
 \caption{Absolute magnitude and derived masses for all confirmed or possible companions detected in our survey. The absolute magnitude refers to the SDSS i-band. If multiple measurements were available, we give the average absolute magnitude. }
 \begin{threeparttable}
  \begin{tabular}{@{}ccc@{}}
  \hline 
 Object 			& abs. mag. $[mag]$	&  mass $[M_\odot]$\\
 \hline
	HD\,2638 		& 8.63 $\pm$ 0.45	&	0.425$^{+0.067}_{-0.095}$	\\
	HAT-P-7			& 10.40$\pm$0.28 	&	0.205$^{+0.026}_{-0.021}$	\\
	HD\,185269\tnote{1}		& 10.10 $\pm$ 0.13 	&	0.232$^{+0.012}_{-0.012}$		\\
	WASP-76			& 6.50 $\pm$ 0.45 	&	0.692$^{+0.074}_{-0.059}$		\\
	HAT-P-32		& 9.00$\pm$0.15 	&	0.340$^{+0.048}_{-0.024}$		\\
	HD\,10697		& 10.75 $\pm$ 0.18	&	0.177$^{+0.013}_{-0.010}$	\\
	HD\,43691\tnote{1}		& 11.06 $\pm$ 0.19 	&	0.160$^{+0.010}_{-0.010}$		\\
	HD\,116029		& 10.7 $\pm$ 1.8 	&	0.18$^{+0.21}_{-0.07}$		\\
	HAT-P-18		& 13.21 $\pm$ 0.17 	&	0.0994$^{+0.0022}_{-0.0016}$		\\
	Kepler-21  		& 8.6$^{+4.2}_{-1.0}$ 	&	0.42$^{+0.14}_{-0.32}$		\\
	Kepler-68		& 10.78 $\pm$ 0.18 	&	0.175$^{+0.013}_{-0.010}$		\\
	Kepler-42		& 15.59 $\pm$ 0.62	&	0.0819$^{+0.0035}_{-0.0029}$		\\
	HD\,197037		& 9.225 $\pm$ 0.066 	&	0.3412$^{+0.0098}_{-0.0477}$		\\
	HD\,217786		& 11.02 $\pm$ 0.13	&	0.1622$^{+0.0071}_{-0.0068}$		\\
        		
\hline\end{tabular}

\begin{tablenotes}
\item[1] we give the unresolved magnitude and the derived mass from that unresolved magnitude
\end{tablenotes}
\end{threeparttable}
\label{tab:masses}
%\end{minipage}
\end{table}

\begin{table}
 %\centering
 %\begin{minipage}{200mm}
 \caption{SPHERE photometric measurements and mass estimates of the resolved components of the binary HD185269\,B. The primary star is saturated in Y and H-band, and thus masses and differential magnitudes could only be calculated in J-band.}
 %\begin{threeparttable}
  \begin{tabular}{@{}cccc@{}}
  \hline 
 Filter					& BB\_Y				&   BB\_J			& BB\_H		\\
 \hline
$\Delta$Ba/Bb $[mag]$	 		&	0.24 $\pm$ 0.11 	&	0.14 $\pm$ 0.12 	& 0.21 $\pm$ 0.04 	\\
$\Delta$A/Ba $[mag]$			&		...		&	6.957 $\pm$ 0.082	&  ...	\\
$\Delta$A/Bb $[mag]$			&		...		&	7.093 $\pm$ 0.088 	&  ...	\\
mass Ba $[M_\odot]$			&	 	...		&	0.165$\pm$0.08 		&  ...	\\
mass Bb $[M_\odot]$      		&	 	...		&	0.154$^{+0.009}_{-0.008}$ 			&  ...	\\
      
\hline\end{tabular}
\label{tab:hd185269-sphere-phot}
%\end{threeparttable}
%\end{minipage}
\end{table}

\begin{table*}
 %\centering
 %\begin{minipage}{200mm}
  \caption{Detection limits of all stars observed in our survey. We give the achievable magnitude difference as well as the corresponding mass limit.}
  \begin{tabular}{@{\extracolsep{4pt}}lcccccccc@{}}
  \hline
							& \multicolumn{2}{c}{0.5 arcsec}  & \multicolumn{2}{c}{1 arcsec}        & \multicolumn{2}{c}{2.5 arcsec}			&		\multicolumn{2}{c}{5 arcsec} \\
	\cline{2-3} \cline{4-5} \cline{6-7} \cline{8-9}
 Star 				& $\Delta$\,mag	 &	M$_{min}$\,[M$_\odot$]			& $\Delta$\,mag 			&		M$_{min}$\,[M$_\odot$]			& $\Delta$\,mag 		&		M$_{min}$\,[M$_\odot$]		& $\Delta$\,mag 	& M$_{min}$\,[M$_\odot$]	\\
		
	 \hline	
		
HD2638		&	3.3	&	 0.402 $^{+0.062}_{-0.091}$ 	&	5.3	&	 0.171 $^{+0.028}_{-0.059}$ 	&	9.0	&	 0.0878 $^{+0.0028}_{-0.043}$ 		&	9.6	&	 0.0843 $^{+0.0027}_{-0.045}$	\\
HD2952		&	3.8	&	...				&	6.4	&	...				&	10.3	&	...					&	10.9	&	...	\\
HD5608		&	2.8	&	...				&	4.1	&	...				&	7.7	&	...					&	9.9	&	...	\\
HD5891		&	2.9	&	 1.22 $^{+0.18}_{-1.20}$ 	&	4.0	&	 1.00 $^{+0.18}_{-0.15}$ 	&	7.5	&	 0.52 $^{+0.12}_{-0.13}$ 		&	9.5	&	 0.243 $^{+0.12}_{-0.076}$	\\
HD8574		&	3.9	&	 0.5564 $^{+0.0068}_{-0.0060}$ 	&	6.9	&	 0.1868 $^{+0.0039}_{-0.0034}$ 	&	10.4	&	 0.09028 $^{+0.00054}_{-0.00043}$ 	&	10.8	&	 0.08800 $^{+0.00034}_{-0.00028}$	\\
HD10697		&	3.8	&	 0.614 $^{+0.019}_{-0.019}$ 	&	6.3	&	 0.275 $^{+0.019}_{-0.019}$ 	&	9.9	&	 0.0994 $^{+0.0021}_{-0.0015}$ 		&	10.1	&	 0.0974 $^{+0.0017}_{-0.0016}$	\\
WASP-76		&	3.3	&	 0.60 $^{+0.11}_{-0.11}$ 	&	5.3	&	 0.34 $^{+0.13}_{-0.14}$ 	&	8.7	&	 0.108 $^{+0.022}_{-0.049}$ 		&	9.1	&	 0.102 $^{+0.016}_{-0.047}$	\\
HAT-P-32	&	3.9	&	 0.554 $^{+0.054}_{-0.028}$ 	&	6.0	&	 0.253 $^{+0.063}_{-0.030}$ 	&	8.4	&	 0.0871 $^{+0.024}_{-0.0074}$ 		&	8.5	&	 0.0826 $^{+0.027}_{-0.0071}$	\\
HD12661		&	3.7	&	 0.498 $^{+0.015}_{-0.017}$ 	&	6.1	&	 0.19823 $^{+0.010}_{-0.0083}$ 	&	9.8	&	 0.0901 $^{+0.0013}_{-0.0011}$ 		&	10.0	&	 0.0881 $^{+0.0011}_{-0.0014}$	\\
HD13189		&	3.7	&	...				&	5.7	&	...				&	9.9	&	...					&	10.3	&	...	\\
HD13908		&	3.8	&	 0.655 $^{+0.034}_{-0.032}$ 	&	6.2	&	 0.329 $^{+0.039}_{-0.037}$ 	&	9.8	&	 0.1046 $^{+0.0037}_{-0.0036}$ 		&	10.1	&	 0.0997 $^{+0.0036}_{-0.0026}$	\\
HD15779		&	2.9	&	...				&	3.9	&	...				&	7.4	&	...					&	9.8	&	...	\\
HD285507	&	3.8	&	 0.189 $^{+0.019}_{-0.016}$ 	&	5.0	&	 0.1266 $^{+0.0085}_{-0.0069}$ 	&	8.6	&	 0.0762 $^{+0.0029}_{-0.0032}$ 		&	9.0	&	 0.0726 $^{+0.0029}_{-0.0035}$	\\
HD290327	&	3.3	&	 0.480 $^{+0.062}_{-0.070}$ 	&	4.5	&	 0.301 $^{+0.076}_{-0.056}$ 	&	8.0	&	 0.1041 $^{+0.0080}_{-0.0060}$ 		&	8.6	&	 0.0965 $^{+0.0054}_{-0.0048}$	\\
HD40979		&	3.4	&	 0.5946 $^{+0.0095}_{-0.0095}$ 	&	5.2	&	 0.347 $^{+0.011}_{-0.011}$ 	&	9.0	&	 0.1025 $^{+0.0011}_{-0.0010}$ 		&	9.8	&	 0.09399 $^{+0.00097}_{-0.00074}$	\\
HD43691		&	3.5	&	 0.652 $^{+0.051}_{-0.047}$ 	&	4.9	&	 0.477 $^{+0.051}_{-0.054}$ 	&	7.7	&	 0.160 $^{+0.021}_{-0.017}$ 		&	8.3	&	 0.133 $^{+0.014}_{-0.011}$	\\
HD45350		&	3.5	&	 0.562 $^{+0.023}_{-0.023}$ 	&	5.1	&	 0.337 $^{+0.029}_{-0.027}$ 	&	8.7	&	 0.1054 $^{+0.0026}_{-0.0025}$ 		&	9.7	&	 0.0942 $^{+0.0019}_{-0.0017}$	\\
Omi Uma		&	3.5	&	$>$1.4				&	5.2	&	 1.017 $^{+0.013}_{-0.011}$ 	&	8.9	&	 0.4966 $^{+0.0080}_{-0.0076}$ 		&	9.6	&	 0.3937 $^{+0.0091}_{-0.0084}$	\\
GJ328		&	3.3	&	...				&	4.4	&	...				&	7.5	&	...					&	8.2	&	...	\\
HD95089		&	3.3	&	 0.914 $^{+0.11}_{-0.098}$ 	&	5.0	&	 0.671 $^{+0.082}_{-0.080}$ 	&	8.6	&	 0.212 $^{+0.067}_{-0.046}$ 		&	9.5	&	 0.155 $^{+0.040}_{-0.027}$	\\
HD96063		&	3.1	&	 0.91 $^{+0.13}_{-0.11}$ 	&	4.2	&	 0.756 $^{+0.11}_{-0.092}$ 	&	7.9	&	 0.282 $^{+0.11}_{-0.075}$ 		&	9.2	&	 0.166 $^{+0.056}_{-0.034}$	\\
HD99706		&	3.5	&	 0.887 $^{+0.077}_{-0.070}$ 	&	5.3	&	 0.652 $^{+0.060}_{-0.059}$ 	&	8.7	&	 0.213 $^{+0.047}_{-0.037}$ 		&	9.3	&	 0.170 $^{+0.034}_{-0.024}$	\\
HD100655	&	3.3	&	 1.150 $^{+0.084}_{-0.072}$ 	&	4.9	&	 0.848 $^{+0.055}_{-0.050}$ 	&	8.9	&	 0.320 $^{+0.050}_{-0.045}$ 		&	9.7	&	 0.225 $^{+0.035}_{-0.030}$	\\
HIP57274	&	3.4	&	 0.285 $^{+0.021}_{-0.017}$ 	&	4.6	&	 0.1740 $^{+0.0099}_{-0.0076}$ 	&	7.4	&	 0.0958 $^{+0.0014}_{-0.0013}$ 		&	7.7	&	 0.0931 $^{+0.0014}_{-0.0014}$	\\
HD102329	&	3.3	&	 0.97 $^{+0.14}_{-0.12}$ 	&	4.8	&	 0.745 $^{+0.11}_{-0.094}$ 	&	8.5	&	 0.263 $^{+0.10}_{-0.071}$ 		&	9.8	&	 0.156 $^{+0.050}_{-0.030}$	\\
HD106270	&	3.1	&	 0.806 $^{+0.051}_{-0.047}$ 	&	4.4	&	 0.643 $^{+0.045}_{-0.043}$ 	&	7.8	&	 0.203 $^{+0.033}_{-0.026}$ 		&	8.8	&	 0.145 $^{+0.018}_{-0.014}$	\\
HD113337	&	3.6	&	 0.671 $^{+0.012}_{-0.013}$ 	&	5.6	&	 0.405 $^{+0.022}_{-0.021}$ 	&	9.3	&	 0.0819 $^{+0.015}_{-0.0076}$ 		&	9.6	&	 0.0752 $^{+0.019}_{-0.0066}$	\\
HD116029	&	4.1	&	 0.751 $^{+0.060}_{-0.054}$ 	&	7.1	&	 0.372 $^{+0.066}_{-0.064}$ 	&	10.2	&	 0.120 $^{+0.014}_{-0.011}$ 		&	10.3	&	 0.1155 $^{+0.013}_{-0.0089}$	\\
HD120084	&	3.4	&	 1.148 $^{+0.035}_{-0.030}$ 	&	4.8	&	 0.896 $^{+0.024}_{-0.021}$ 	&	8.8	&	 0.356 $^{+0.020}_{-0.019}$ 		&	9.8	&	 0.227 $^{+0.013}_{-0.012}$	\\
Beta UMi	&	3.5	&	$>$1.3				&	4.6	&	$>$1.3				&	8.0	&	 0.7522 $^{+0.010}_{-0.0084}$ 		&	9.6	&	 0.5536 $^{+0.0073}_{-0.0064}$	\\
HD131496	&	4.0	&	 0.741 $^{+0.062}_{-0.057}$ 	&	6.9	&	 0.370 $^{+0.071}_{-0.069}$ 	&	10.1	&	 0.117 $^{+0.014}_{-0.010}$ 		&	10.3	&	 0.1115 $^{+0.014}_{-0.0072}$	\\
HD136726	&	4.2	&	 1.240 $^{+0.071}_{-0.045}$ 	&	7.0	&	 0.777 $^{+0.021}_{-0.020}$ 	&	10.3	&	 0.349 $^{+0.021}_{-0.020}$ 		&	10.5	&	 0.320 $^{+0.022}_{-0.020}$	\\
11 UMi		&	3.3	&	$>$1.2				&	4.7	&	 1.120 $^{+0.044}_{-0.040}$ 	&	8.6	&	 0.589 $^{+0.017}_{-0.016}$ 		&	9.7	&	 0.438 $^{+0.021}_{-0.020}$	\\
HD136512	&	4.1	&	 1.008 $^{+0.042}_{-0.038}$ 	&	7.0	&	 0.590 $^{+0.025}_{-0.024}$ 	&	10.5	&	 0.170 $^{+0.013}_{-0.011}$ 		&	10.7	&	 0.160 $^{+0.011}_{-0.011}$	\\
Omi CrB		&	3.5	&	 1.040 $^{+0.074}_{-0.058}$ 	&	4.8	&	 0.852 $^{+0.043}_{-0.037}$ 	&	8.5	&	 0.373 $^{+0.032}_{-0.031}$ 		&	10.1	&	 0.192 $^{+0.017}_{-0.014}$	\\
HD139357	&	4.4	&	 1.03 $^{+0.19}_{-1.00}$ 	&	7.3	&	 0.674 $^{+0.068}_{-0.065}$ 	&	10.2	&	 0.291 $^{+0.079}_{-0.058}$ 		&	10.3	&	 0.283 $^{+0.076}_{-0.056}$	\\
HD145457	&	4.5	&	 0.896 $^{+0.053}_{-0.049}$ 	&	7.2	&	 0.536 $^{+0.040}_{-0.041}$ 	&	10.3	&	 0.170 $^{+0.021}_{-0.017}$ 		&	10.4	&	 0.163 $^{+0.019}_{-0.016}$	\\
HD152581	&	3.9	&	 0.79 $^{+0.15}_{-0.13}$ 	&	6.8	&	 0.46 $^{+0.14}_{-0.16}$ 	&	9.7	&	 0.146 $^{+0.069}_{-0.036}$ 		&	9.8	&	 0.142 $^{+0.064}_{-0.033}$	\\
HAT-P-18	&	4.1	&	 0.222 $^{+0.028}_{-0.023}$ 	&	6.6	&	 0.1067 $^{+0.0045}_{-0.0037}$ 	&	7.9	&	 0.0923 $^{+0.0028}_{-0.0025}$ 		&	8.0	&	 0.0915 $^{+0.0028}_{-0.0021}$	\\
HD156279	&	4.0	&	 0.389 $^{+0.015}_{-0.013}$ 	&	6.5	&	 0.1478 $^{+0.0039}_{-0.0035}$ 	&	10.0	&	 0.08598 $^{+0.00050}_{-0.00050}$ 	&	10.3	&	 0.08471 $^{+0.00050}_{-0.00050}$	\\
HD163607	&	3.7	&	 0.609 $^{+0.020}_{-0.018}$ 	&	6.2	&	 0.271 $^{+0.019}_{-0.018}$ 	&	9.6	&	 0.1019 $^{+0.0023}_{-0.0022}$ 		&	9.8	&	 0.0991 $^{+0.0020}_{-0.0015}$	\\
HD163917	&	3.8	&	 1.289 $^{+0.021}_{-0.019}$ 	&	6.6	&	 0.7652 $^{+0.0095}_{-0.0095}$ 	&	10.5	&	 0.2481 $^{+0.0075}_{-0.0072}$ 		&	11.2	&	 0.1843 $^{+0.0050}_{-0.0048}$	\\
HIP91258	&	3.4	&	 0.451 $^{+0.024}_{-0.044}$ 	&	5.0	&	 0.234 $^{+0.016}_{-0.055}$ 	&	8.7	&	 0.0939 $^{+0.0016}_{-0.040}$ 		&	9.8	&	 0.08582 $^{+0.00099}_{-0.042}$	\\
Kepler-37	&	3.5	&	 0.40 $^{+0.33}_{-0.25}$ 	&	5.5	&	 0.177 $^{+0.30}_{-0.078}$ 	&	9.0	&	 0.089 $^{+0.040}_{-0.013}$ 		&	9.5	&	 0.086 $^{+0.027}_{-0.011}$	\\
Kepler-21	&	3.7	&	 0.691 $^{+0.043}_{-0.043}$ 	&	6.0	&	 0.395 $^{+0.052}_{-0.051}$ 	&	9.0	&	 0.129 $^{+0.014}_{-0.010}$ 		&	9.2	&	 0.124 $^{+0.012}_{-0.010}$	\\
HD180314	&	4.4	&	 0.924 $^{+0.054}_{-0.048}$ 	&	7.2	&	 0.544 $^{+0.035}_{-0.035}$ 	&	10.2	&	 0.176 $^{+0.020}_{-0.019}$ 		&	10.4	&	 0.168 $^{+0.018}_{-0.017}$	\\
Kepler-63	&	3.7	&	 0.424 $^{+0.059}_{-0.063}$ 	&	5.8	&	 0.171 $^{+0.034}_{-0.032}$ 	&	8.0	&	 0.081 $^{+0.014}_{-0.011}$ 		&	8.1	&	 0.081 $^{+0.014}_{-0.012}$	\\
Kepler-68	&	4.0	&	 0.485 $^{+0.049}_{-0.054}$ 	&	6.6	&	 0.175 $^{+0.027}_{-0.020}$ 	&	9.0	&	 0.1003 $^{+0.0052}_{-0.0038}$ 		&	9.1	&	 0.0985 $^{+0.0046}_{-0.0037}$	\\
Kepler-42	&	4.8	&	 0.0794 $^{+0.0045}_{-0.0040}$ 	&	6.4	&	 0.0760 $^{+0.0020}_{-0.0061}$ 	&	6.6	&	 0.0756 $^{+0.0020}_{-0.0062}$ 		&	6.6	&	 0.0757 $^{+0.0020}_{-0.0062}$	\\
HAT-P-7		&	3.6	&	 0.704 $^{+0.082}_{-0.079}$ 	&	5.9	&	 0.414 $^{+0.090}_{-0.090}$ 	&	8.8	&	 0.136 $^{+0.029}_{-0.019}$ 		&	9.0	&	 0.127 $^{+0.025}_{-0.018}$	\\
HD188015	&	4.5	&	 0.424 $^{+0.040}_{-0.040}$ 	&	7.3	&	 0.143 $^{+0.012}_{-0.010}$ 	&	10.3	&	 0.0881 $^{+0.0015}_{-0.0015}$ 		&	10.5	&	 0.0870 $^{+0.0015}_{-0.0015}$	\\
HD190360	&	3.7	&	 0.4873 $^{+0.0088}_{-0.0062}$ 	&	6.2	&	 0.1886 $^{+0.0042}_{-0.0031}$ 	&	9.9	&	 0.08902 $^{+0.00031}_{-0.00028}$ 	&	10.3	&	 0.08710 $^{+0.00031}_{-0.00025}$	\\
HD197037	&	3.8	&	 0.5255 $^{+0.0087}_{-0.024}$ 	&	6.2	&	 0.212 $^{+0.011}_{-0.043}$ 	&	10.0	&	 0.077 $^{+0.011}_{-0.025}$ 		&	10.4	&	 0.072 $^{+0.012}_{-0.024}$	\\
HD206610	&	3.2	&	 0.97 $^{+0.21}_{-0.17}$ 	&	4.8	&	 0.74 $^{+0.16}_{-0.14}$ 	&	8.6	&	 0.240 $^{+0.15}_{-0.084}$ 		&	9.6	&	 0.161 $^{+0.088}_{-0.043}$	\\
HD208527	&	4.0	&	$>$1.4				&	6.9	&	 1.16 $^{+0.24}_{-1.2}$ 	&	10.1	&	 0.66 $^{+0.12}_{-0.11}$ 		&	10.3	&	 0.63 $^{+0.12}_{-0.11}$	\\
HD210277	&	3.3	&	 0.5234 $^{+0.0099}_{-0.0089}$ 	&	5.7	&	 0.2152 $^{+0.0070}_{-0.0064}$ 	&	9.6	&	 0.09015 $^{+0.00069}_{-0.00042}$ 	&	10.2	&	 0.08645 $^{+0.00040}_{-0.00036}$	\\
HD217786	&	4.0	&	 0.532 $^{+0.027}_{-0.025}$ 	&	5.4	&	 0.322 $^{+0.031}_{-0.029}$ 	&	9.5	&	 0.0984 $^{+0.0023}_{-0.0020}$ 		&	9.8	&	 0.0948 $^{+0.0020}_{-0.0020}$	\\
HD240210	&	3.6	&	 0.85 $^{+0.35}_{-0.85}$ 	&	5.8	&	 0.61 $^{+0.22}_{-0.25}$ 	&	9.7	&	 0.159 $^{+0.17}_{-0.056}$ 		&	10.4	&	 0.129 $^{+0.12}_{-0.034}$	\\
HD219828	&	4.1	&	 0.559 $^{+0.035}_{-0.035}$ 	&	7.0	&	 0.194 $^{+0.024}_{-0.020}$ 	&	9.8	&	 0.0984 $^{+0.0033}_{-0.0028}$ 		&	9.9	&	 0.0974 $^{+0.0029}_{-0.0027}$	\\
HD220074	&	3.7	&	$>$1.2				&	6.1	&	 1.16 $^{+0.24}_{-1.2}$ 	&	9.8	&	 0.64 $^{+0.11}_{-0.10}$ 		&	10.2	&	 0.58 $^{+0.11}_{-0.10}$	\\
HD222155	&	3.4	&	 0.654 $^{+0.019}_{-0.019}$ 	&	4.9	&	 0.468 $^{+0.023}_{-0.021}$ 	&	8.8	&	 0.1177 $^{+0.0045}_{-0.0043}$ 		&	9.6	&	 0.1026 $^{+0.0023}_{-0.0021}$	\\

\hline\end{tabular}
\label{tab:detection-output}
%\end{minipage}
\end{table*}

\section{Discussion of the new bound stellar companions}

\subsection{Kepler-21}

Kepler-21 (also known as HD\,179070, KOI-975, KIC 3632418) is the brightest star in the original Kepler sample. \cite{2012ApJ...746..123H} found a transiting planet of approximately 1.6 times the size of the earth in a $\sim$2.8 day orbit around this star. According to them, the planet has an upper mass limit of 10.5 Earth masses and is moving on a circular orbit. They also carried out high resolution adaptive optics imaging of the host star with the Keck telescope in the near infrared. In these images taken on 22-02-2011 they detected a faint source with a separation of 0.75\,arcsec at a position angle of 129\,deg. This source is identical to the source that we detected with AstraLux in our 2013 and 2014 observations and that emerged as new co-moving low-mass stellar companion. 
We introduce this companion here as a new discovery, because \cite{2012ApJ...746..123H} exclude the possibility that the source is physically associated to the host star based on its J-K$_s$ color. They argue that the color of the companion is either consistent with a late M dwarf which should then be located at $\sim$15\,pc or with a M0 giant, which would be located in an approximate distance of 10\,kpc. Since Kepler-21 is located at approximately 112\,pc, the two sources should then not be associated. However, our own astrometric measurements in 2013 and 2014 show clearly that the source is co-moving with Kepler-21. In fact, also the astrometric position given by \cite{2012ApJ...746..123H} in their 2011 Keck measurement is perfectly consistent with a co-moving object. If the object was indeed a background giant in some kpc distance, we would have expected a position angle of 119.6\,deg at the time of the 2011 meassurement. Unfortunately \cite{2012ApJ...746..123H} do not provide uncertainties for their astrometric meassurements. However, a deviation of almost 10\,deg seems very unlikely.
To get an estimate of the likelihood to detect a background or foreground object within 0.77\,arcsec around Kepler-21 we followed the approach by \cite{2014A&A...566A.103L}. They give the probability to find a physically unrelated source within a certain distance of a star with 

\begin{equation}
P(r,b,m_{\odot},\Delta m_{max}) = \pi r^2 \rho (b,m_{\odot},\Delta m_{max}) \ ,  
\label{prob} 
\end{equation}

wherein $r$ is the separation from the star, $b$ is the galactic latitude, $m_{\odot}$ is the apparent magnitude of the star in the observed filter, $\Delta m_{max}$ is the maximum achieved contrast within the separation $r$, and $\rho$ is the stellar density. 
To estimate the stellar density as a function of the galactic latitude and the achieved magnitude limit, we utilize the TRILEGAL\footnote{http://stev.oapd.inaf.it/cgi-bin/trilegal\_1.6} population synthesis code by \cite{2005A&A...436..895G}. We choose the default parameters for the different parts of the Galaxy and the lognormal initial mass function of \cite{2001ApJ...554.1274C}.
We find that in an area of 1\,deg$^2$ around Kepler-21 we should be able to detect 452 stars with a limiting magnitude of 12.86\,mag in SDSS i-band. The limiting magnitude is the value that we are computing as described in section \ref{sec: detect} for a separation of 0.8\,arcsec.
This yields a stellar density $\rho$ of 3.5$\cdot10^{-5}$ sources per arcsec$^2$. Putting this into equation \ref{prob} we find a probability of 6.5$\cdot10^{-5}$ to detect an unrelated background or foreground source within 0.77\,arcsec of Kepler-21.
We thus conclude that, given our astrometry, the most likely explanation is indeed that the companion candidate is physically bound to Kepler-21.\\
Kepler-21\,B is located at a projected separation of only 87\,au, which might indicate that it should have had a strong influence on the planet formation process. One possible scenario might be that the stellar companion excited high eccentricities in the forming planet causing close encounters with the primary star. The eccentricity could have then been damped by tidal heating which would have left the companion on a very short periodic circular orbit. Such scenarios have been suggested to occur in multiple planetary systems, where multiple objects interact dynamically, e.g. by \cite{1996Sci...274..954R}. Given that the system is evolved ($\sim$3.6\,Gyr), it is consistent that we would now observe the end product of this interaction.     

\subsection{Kepler-68}

The star Kepler-68 hosts three known planets detected via transit and radial velocity observations by \cite{2013ApJ...766...40G}. The innermost two of these planets have orbit periods in the order of days and masses in the order of several Earth masses and were detected in transit, while the outer planet d was found in radial velocity data and has a much longer orbit period of $\sim$1.6\,yr (semi-major axis of 1.4\,au) and higher mass (m$\cdot$sin(i) = 0.95\,M$_{Jup}$). The inner planets appear to be on circular orbits while the outer planet exhibits an orbit eccentricity of 0.18. \\
The newly discovered stellar component Kepler-68\,B is located at a projected separation of 1485\,au. Due to this large separation, the expected period of Kozai-Lidov type resonances is in the order of several Gyr. It seems thus unlikely that the stellar component has a major influence on the dynamics of the inner system via this mechanism. It remains unclear if such a widely separated outer stellar component has a strong influence on the circumstellar disk in the planet formation phase. 
Following the argument of \cite{2012ApJ...745...19K}, who studied the occurence rate of circumstellar disks in young binary systems, a major influence of the secondary stellar component is only expected for separations of up to 40\,au. If this observational result holds true, then Kepler-68\,B is too widely separated to have influenced the circumstellar disk around Kepler-68\,A. However, it is in principle also possible that the B component is on a very eccentric orbit. If this is the case then close interactions with the inner planets or the planet forming disk might have happend. In the case of a very eccentric orbit, we would expect to find the stellar companion at a wide separation since it spends the majority of the time there. Further high precision astrometric monitoring combined with statistical orbit analysis might shed some light on the orbit of the B component.  \\
Since the source that we now identified as Kepler-68\,B was detected in 2MASS, it was included in the Kepler input catalogue. With its large separation of ~11arcsec the new stellar component is not within the "classical" Kepler PSF. However, Kepler-68 is strongly saturated and shows bleeding. Therefore, the changes in flux can only be seen at the end of bleed columns. 
 \cite{2013ApJ...766...40G} showed that Kepler-68\,B is located almost precisely in the column direction from Kepler-68. 
In most of the observing quarters of Kepler, the bleeding encompasses Kepler-68\,B and, hence, has to be taken into account for the transit measurements. 
If the light contribution of B is not considered, systematic errors in the system parameters will arise without changing the quality of the transit fit. 
From the measured magnitude difference of ~6.6\,mag (see Tab.~\ref{tab:measurements}) we calculated the amount of contaminating light. 
If B is a binary that exhibits total eclipses, the transit depth measured by Kepler would be ~2300\,ppm. 
This is much higher than the  detected transit depth of 346\,ppm and 55\,ppm for Kepler-68\,b and Kepler-68\,c, respectively (\citealt{2013ApJ...766...40G}). 
Therefore a partial or grazing eclipse of B could produce a transit signal. 
However, as shown before by e.g. \cite{2011ApJ...732L..24L} or \cite{2011ApJS..197....5F}, it is very unlikely that an eclipsing background object can mimic a multiple planetary system. 
Furthermore \cite{2013ApJ...766...40G} showed that in Quarter 9 Kepler-68 as well as Kepler-68\,B are located between columns in such a way that the bleeding terminates before reaching the latter. 
In this way they proved that Kepler-68\,B cannot be the source of the transit signal. Finally, by applying the BLENDER procedure (\citealt{2004ApJ...614..979T}, \citealt{2011ApJS..197....5F}), \cite{2013ApJ...766...40G} could rule out all false positive scenarios involving eclipsing binaries and validate Kapler-68\,b,c as planets.

\subsection{HD\,197037}

\begin{figure}

\includegraphics[scale=0.4]{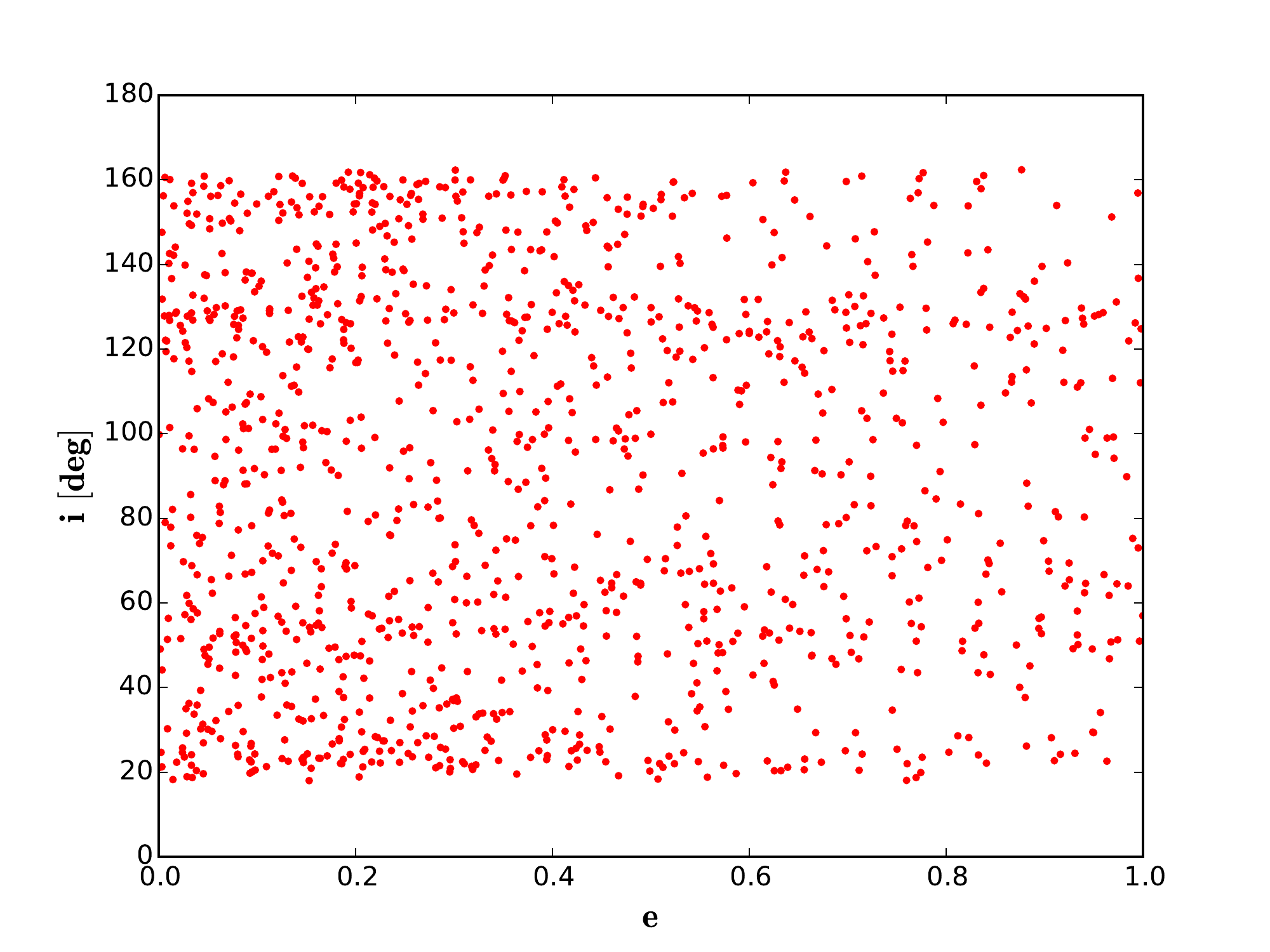}

\caption[]{Inclination and eccentricity distribution for possible orbits of HD197037\,B that induce a linear trend in the radial velocity of HD197037\,A as measured by \cite{2012ApJ...749...39R}. Shown are 1217 out of 15000 randomly generated bound keplerian orbits that include the current position of HD197037\,B and that match the total system mass.} 
\label{hd197037-orbit}
\end{figure}

HD\,197037 hosts an m$\cdot$sin(i) = 0.79 M$_{Jup}$ planet on a $\sim$2.8\,yr period (semi-major axis of 2.1\,au), discovered by \cite{2012ApJ...749...39R}. They found that their best fitting orbit solution for the planet exhibits an eccentricity of 0.22. They also note that they include in their model a linear trend in radial velocity with a slope of -1.87$\pm$0.3\,ms$^{-1}$yr$^{-1}$, which could be attributed to a long period planet of 0.7\,M$_{Jup}$ and a period of $\sim$12\,yr, or possibly a more distant stellar companion of which they find no further evidence.\\
To determine if the newly detected stellar companion HD\,197037\,B can be responsible for this linear trend in the radial velocity, we performed a dedicated Monte Carlo simulation. We fixed the system mass to the combined mass of both stellar components, i.e. 0.34\,M$_\odot$ for B and 1.11\,M$_\odot$ for A (\citealt{2012ApJ...749...39R}). We then generated random bound Keplerian orbits which are compatible with our astrometric measurement of HD\,197037\,B. To somewhat narrow the wide parameter space, we restricted our simulation to orbits with a semi-major axis between 3\,arcsec and 6\,arcsec and times of periastron passage within 2000\,yr from our astrometric epoch.
We created a total of 15000 such orbits. We then checked which of these orbits would introduce a slope as measured by \cite{2012ApJ...749...39R} in their measurement period between the beginning of 2001 and 2012. Out of the 15000 randomely generated orbits, 1217 orbits fulfill this criterion. In Fig.~\ref{hd197037-orbit}, we show the eccentricities and orbit inclinations of all these orbits.
We find that there is no strongly preferred region of the parameter space for an orbit of the B component to produce the measured radial velocity slope. In particular we find orbits for the full range of possible eccentricities. The range of possible inclinations is constrained only by the photometric mass estimate of the B component, i.e. the orbit needs to have a minimum inclination of $\sim$18\,deg to produce the radial velocity signal. From our imaging epochs we can not yet constrain the orbit of the B component, i.e. it is in principle possible that the B component is in a face-on or close to face-on orbit configuration. 
However given the large range of orbit solutions of the B component that reproduce the measured linear radial velocity trend, we find it likely that this trend is indeed caused by the stellar B component and not by an additional long period planet.\\
The non-circular orbit of the existing extrasolar planet around the A component might also be well explained by the new stellar companion if they are cought in mutual Kozai-Lidov type resonances. Given the potential very young age of the system of 0.3$\pm$0.3\,Gyr (\citealt{2015AandA...575A..18B}), it may also be possible that the stellar B component was not originally a part of the system but was just caught as the result of a stellar flyby in more recent times. This would then have disrupted the original circular orbit of the planet. However, since HD197037 is not a known member of a star forming region or young moving group, such an event would seem rather unlikely.

\subsection{HD\,217786}

\cite{2011A&A...527A..63M} discovered a long period ($\sim$3.6\,yr) planet or brown dwarf with a minimum mass of 13\,M$_{Jup}$ around HD\,217786 via radial velocity measurements. They found that the best fitting orbit solution of the object is very eccentric with e = 0.4$\pm$0.05. They do not see long term radial velocity trends in their data.\\
The new stellar companion HD\,217786\,B is located at a projected separation of 155\,au. To explore whether the large eccentricity of the planetary companion could be caused by Kozai-Lidov resonances with the stellar companion, we calculated the period of possible Kozai cycles. For this purpose we used the formula provided in \cite{2005ApJ...627.1001T}. We assumed that the semi-major axis of the orbit of the stellar companion is equal to its projected separation and that the orbit is circular. We get a period of $\sim$ 6.2\,Myr. This period can be approximately an order of magnitude shorter if the stellar companion is on a significantly eccentric orbit itself. Given the large system age of $\sim$6.5\,Gyr, more than a thousand Kozai cycles could have been complete in principle. It is thus conceivable that the eccentricity of the planetary companion is indeed caused by such interactions with the newly discovered stellar companion.
However, this is just one possible scenario to explain the eccentricity of the planet and it strongly depends on the actual orbit of the new stellar companion.

\section{Summary}

We searched for stellar companions around 63 stars known to harbor extrasolar planets using AstraLux at the Calar Alto observatory. We found previously unknown faint companion candidates within the field of view of our observations around 11 of the observed systems. Of these companion candidates, four, namely Kepler-21\,B, Kepler-68\,B, HD\,197037\,B and HD\,217786\,B, emerged as co-moving, and thus in all likelihood gravitationally bound, companions. The candidates detected around HD188015 and Kepler-37 are more consistent with background objects. For the remaining 5 systems follow-up lucky imaging observations must still be performed to determine the status of the objects, i.e. if they are co-moving with the exoplanet host star. The candidate found next to HD\,43691 might be of special interest since it may be a low-mass binary itself.\\
We also present new photometric and astrometric measurements for the previously known companions to the HD\,2638, HAT-P-7, HD\,185269, WASP-76 and HAT-P-32 systems. Our SPHERE observations of HD\,185269\,B showed that the companion is actually a very low-mass binary itself, making the system one of only 17 triple systems known to harbor extrasolar planets. Continued astrometric monitoring within the next decade will allow us to determine the dynamical mass of the binary companion.\\
We note that the previously detected companion candidate to WASP-76 (\citealt{2015AandA...579A.129W}) is more consistent with a background source given our new astrometric measurement, however no final conclusion could be drawn due to the short time baseline between the two observational epochs.\\
Including the first part of our survey presented in \cite{2012MNRAS.421.2498G} we have now studied the multiplicity of 128 known exoplanet systems. In this sample we found so far 7 new confirmed binary systems. This includes the new systems reported by us in this work and in \cite{2012MNRAS.421.2498G}, as well as all systems that were first reported in other studies, but that were unknown at the time of our first epoch observation. 
This yields a multiplicity rate of only 5.5\% in our sample. This is much lower than previous values reported by \cite{2012A&A...542A..92R} (12\%), \cite{2014MNRAS.439.1063M} (13\%) or \cite{2015MNRAS.450.3127M} (9\%). If most of the unconfirmed companions that we report in this study turn out to be bound companions, the multiplicity rate of our study would increase to 9-10\%, which would be in better agreement with previous results. 
One contributing factor to our lower multiplicity rate might be that the majority of our sample is comprised of planetary systems found via the radial velocity method. Radial velocity surveys routinely exclude known binary systems from their target sample. Thus they introduce an inherent bias towards single star systems. However, the same was in principle true for the studies by \cite{2014MNRAS.439.1063M} and \cite{2015MNRAS.450.3127M}.\\
If the low multiplicity rate that we recover is indeed caused by a bias introduced by radial velocity surveys, then it would be expected that a higher stellar multiplicity rate is found for transiting planets. \cite{2015ApJ...813..130W} present the results of an adaptive optics imaging search around 138 Kepler planet hosts. They find a stellar multiplicity rate of 8.0$\pm$4.0\,\% for multi-planet systems and 6.4$\pm$5.8\,\% for single-planet systems and stellar companions with semi-major axes between 100\,au and 2000\,au.
These values are in principle consistent with the stellar multiplicity rate of 5.5\,\% that we find, which might indicate that the selection bias of radial velocity surveys has no significant influence on our result.
However, from a statistical point of view, considering simple random sampling, our sample size is too small for accurate predictions. If we assume a confidence level of 95\%, then our estimated level of accuracy for a population size of 1200 exoplanet systems is only 8.2\%. Given that our sample is definitely biased towards single star systems, our actual level of accuracy will be worse than this estimate.
To get a reliable estimate of the stellar multiplicity of exoplanet systems with a margin of error on the 5\% level, a random sample size of 291 systems is neccessary considering the known population of $\sim$1200 confirmed systems. If we consider a much larger population, i.e. all planetary systems in the Galaxy, then a larger random sample size of 385 systems is needed. These are again lower limits considering the potential biases introduced by exoplanet surveys.
We are continuing our mutiplicity survey in order to provide a homogeneous observation base for statistical analysis.

%__________________________________________________________________

\section*{Acknowledgments}

We would like to thank the very helpful staff at the Calar Alto Observatory for their assistance in carrying out our observations. In particular we would like to thank S. Pedraz, G. Bergond and F. Hoyo for organizing and carrying out service mode observations for us.
CG and MM thank DFG for support in projects MU2695/13-1, MU2695/14-1, MU2695/15-1, MU2695/16-1, MU2695/18-1, MU2695/20-1, MU2695/22-1, MU2695/23-1. HA acknowledges support from the Millennium Science Initiative (Chilean Ministry of Economy), through grant "Nucleus P10-022-F" and from FONDECYT grant 3150643. MS thanks Piezosystem Jena for financial support.
SPHERE is an instrument designed and built by a consortium consisting of IPAG (Grenoble, France), MPIA (Heidelberg, Germany), LAM (Marseille, France), LESIA (Paris, France), Laboratoire Lagrange (Nice, France), INAF - Osservatorio di Padova (Italy), Observatoire de Geneve (Switzerland), ETH Zurich (Switzerland), NOVA (Netherlands), ONERA (France) and ASTRON (Netherlands) in collaboration with ESO. SPHERE was funded by ESO, with additional contributions from CNRS (France), MPIA (Germany), INAF (Italy), FINES (Switzerland) and NOVA (Netherlands). SPHERE also received funding from the European Commission Sixth and Seventh Framework Programmes as part of the Optical Infrared Coordination Network for Astronomy (OPTICON) under grant number RII3-Ct-2004-001566 for FP6 (2004-2008), grant number 226604 for FP7 (2009-2012) and grant number 312430 for FP7 (2013-2016). 
This research has made use of the SIMBAD database as well as the VizieR catalogue access tool, operated at CDS, Strasbourg, France. This research has made use of NASA's Astrophysics Data System Bibliographic Services. 
Finally CG would like to thank Donna Keeley for language editing of the manuscript.

\bibliographystyle{mnras}
\bibliography{myBib}

\label{lastpage}

\end{document}